\documentclass{article}
\usepackage{cite}
\usepackage{float}
\usepackage{hyperref}
\usepackage{amsmath,amssymb,amsfonts}
\usepackage{algorithm}
\usepackage{setspace}
\usepackage[table]{xcolor}
\usepackage{colortbl}
\usepackage{tabularx}  
\usepackage{calc} 
\usepackage{subcaption}
\usepackage{times}
\usepackage{epsfig}
\usepackage{makecell}
\usepackage{multirow}
\usepackage{url}            
\usepackage{booktabs}       
\usepackage{amsfonts}       
\usepackage{nicefrac}       
\usepackage{microtype}      
\usepackage{amsmath}
\usepackage{bbding}
\usepackage{float}
\usepackage{caption}
\usepackage{subcaption}
\usepackage{wrapfig}

\usepackage{graphicx}
\usepackage{textcomp}
\usepackage{xcolor}
\usepackage{listings}
\usepackage{adjustbox}
\usepackage{helvet}
\usepackage[accepted]{mlsys2025}
\usepackage{pifont}

\usepackage{eso-pic}
\usepackage{xparse}

\newlength{\badgewidth}
\newlength{\badgegap}
\newcommand{\badgeList}{}

\NewDocumentCommand{\addTopRightBadge}{O{} m}{%
\gappto{\badgeList}{\href{#1}{\includegraphics[width=\badgewidth]{#2}}\hspace{\badgegap}}%
}

\newcommand{\placeTopRightBadges}{%
\AddToShipoutPictureBG*{%
\put(\LenToUnit{\paperwidth - 1.5cm - \badgewidth},\LenToUnit{\paperheight - 2cm}){%
\makebox[0pt][r]{\badgeList}%
}%
}%
}

\addTopRightBadge{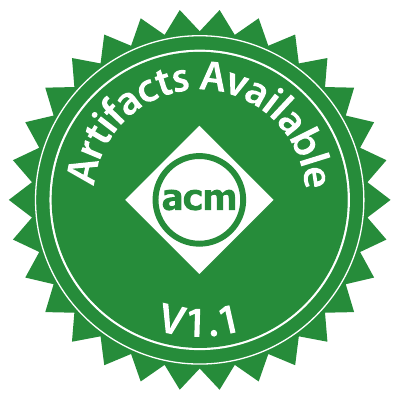}
\addTopRightBadge{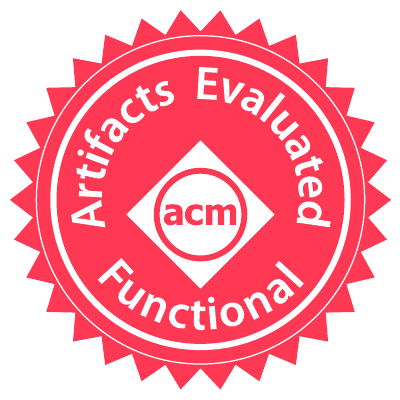}

\placeTopRightBadges
    
\mlsystitlerunning{Training Ultra Long Context Language Model with Fully Pipelined Distributed Transformer}

\begin{document}
\twocolumn[
\mlsystitle{Training Ultra Long Context Language Model with Fully Pipelined Distributed Transformer}


\begin{mlsysauthorlist}
\mlsysauthor{Jinghan Yao\textsuperscript{*}}{1}
\mlsysauthor{Sam Ade Jacobs}{2}
\mlsysauthor{Masahiro Tanaka}{2}
\mlsysauthor{Olatunji Ruwase}{2}
\mlsysauthor{Hari Subramoni}{1}
\mlsysauthor{Dhabaleswar K. (DK) Panda}{1}
\end{mlsysauthorlist}

\mlsysaffiliation{1}{The Ohio State University}
\mlsysaffiliation{2}{Microsoft Inc}

\mlsyskeywords{Machine Learning, MLSys}

\vskip 0.3in

\begin{abstract}

Large Language Models (LLMs) with long context capabilities are integral to complex tasks in natural language processing and computational biology, such as text generation and protein sequence analysis. However, training LLMs directly on extremely long contexts demands considerable GPU resources and increased memory, leading to higher costs and greater complexity. Alternative approaches that introduce long context capabilities via downstream finetuning or adaptations impose significant design limitations. In this paper, we propose Fully Pipelined Distributed Transformer (FPDT) for efficiently training long-context LLMs with outstanding hardware efficiency.
For GPT and Llama models, we achieve a 16x increase in sequence length that can be trained on the same hardware compared to current state-of-the-art solutions. With our dedicated sequence chunk pipeline design, we can now train 8B LLM with 2 million sequence length on only 4 GPUs, while also maintaining over $55\%$ of MFU.
Our proposed FPDT is agnostic to existing training techniques and is proven to work efficiently across different LLM models. \href{https://www.deepspeed.ai/tutorials/ulysses-offload/}{Code is available.}

\end{abstract}
]

\mlsyscorrespondingauthor{Jinghan Yao}{yao.877@osu.edu}
\mlsyscorrespondingauthor{Sam Ade Jacobs}{samjacobs@microsoft.com}

\printAffiliationsAndNotice{\textsuperscript{*}Work done while intern at Microsoft.}

\section{Introduction}

The rapid advancement of large language models (LLMs) has significantly impacted natural language processing (NLP), driving improvements across a wide range of applications. As LLMs like GPT-4, Claude, and Gemini become increasingly capable of processing regular prompts,
there is a growing demand to extend their context windows to accommodate longer input sequences. This capability is crucial for a variety of applications, including comprehensive document analysis, where models must process entire legal documents or scientific papers~\cite{xiong2023effective, peng2023yarn}; long-form content generation, such as writing books or detailed reports; maintaining coherent and contextually relevant long-term dialogues in conversational AI~\cite{beltagy2020longformer, munkhdalai2024leave, mosaicml2023mpt7b, touvron2023llama}; and handling complex multi-step reasoning tasks in fields like healthcare~\cite{zvyagin2023genslms, li2022clinical, gao2021limitations}, climate~\cite{nguyen2023climax}, and finance~\cite{yang2023fingpt, li2023large, kim2024bloated, eisfeldt2023generative, kim2023transcripts}.

However, LLM training is typically constrained to relatively short context lengths, such as 8K or 32K tokens. Currently, most LLMs utilize rotary position embedding~\cite{su2024roformer} (RoPE) to encode input tokens. Despite its improved efficiency and extrapolation capability compared to regular position embeddings~\cite{vaswani2017attention}, to accommodate much longer prompts during inference, RoPE often requires aggressive rescale and map. These adjustments struggle to properly adapt models to a longer context before the model's performance deteriorates, leading to a collapse in the quality of its outputs.
This presents a critical challenge: the necessity to train LLMs originally on the desired long context lengths to ensure robust performance across varying applications.

\begin{figure}[t]
    \centering
    \includegraphics[width=0.47\textwidth]{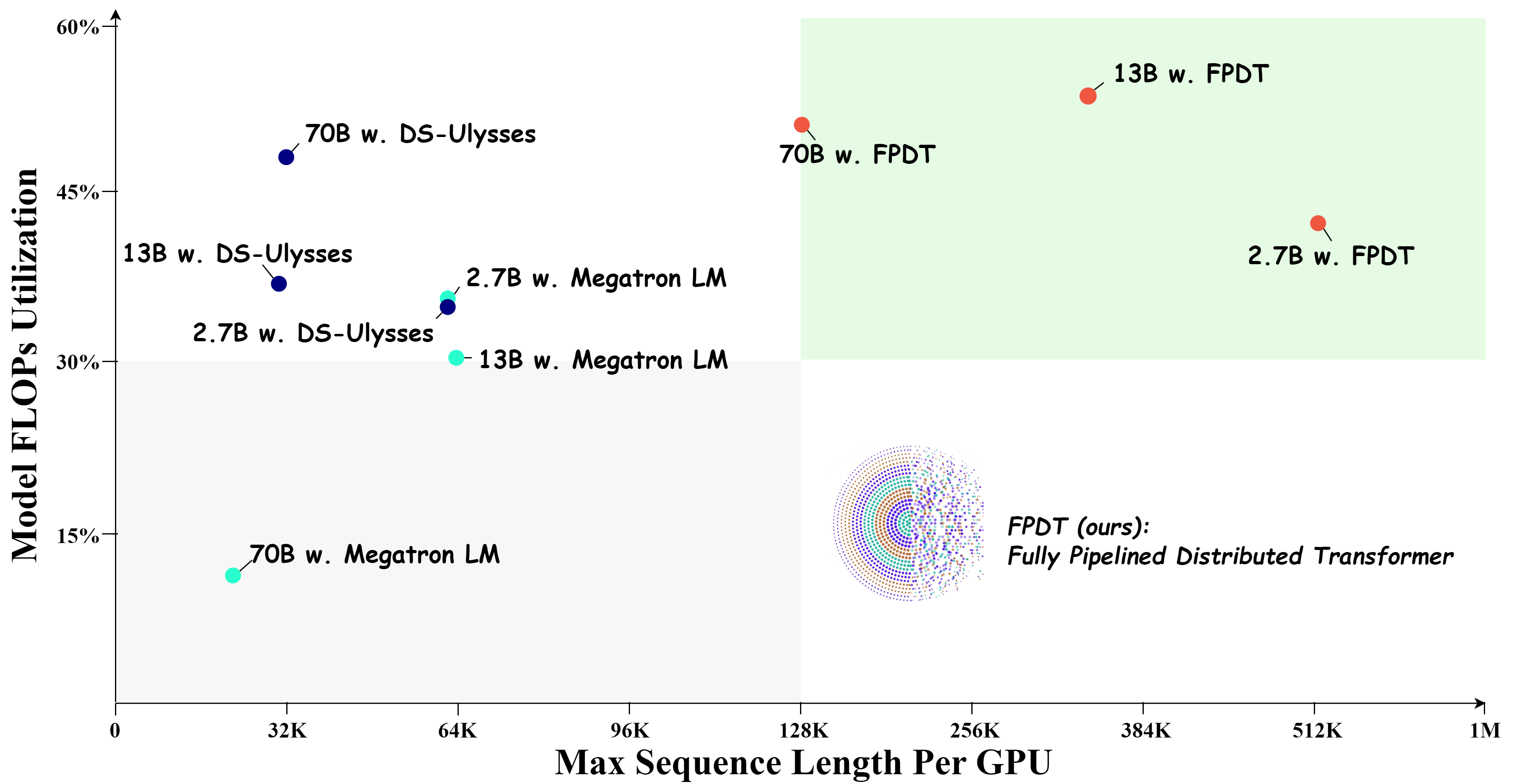}
    \caption{Comparison of end-to-end training Model FLOPs Utilization (MFU) and maximum context length per GPU supported. We show 3 model sizes, i.e. 2.7B, 13B, and 70B.}
    \label{fig:avg_time}
    \vspace{-2em}
\end{figure}

\begin{table*}[ht]
    \centering
    {\fontfamily{ppl}\selectfont
        \begin{tabular}{c c|c|c|c|c|c|c|c}
            \bottomrule[1pt]
            \multirow{3}{*}{\textbf{Model size}} & \multicolumn{8}{c}{\textbf{Hardware configuration}} \\ 
            \cline{2-9}
            & \multicolumn{4}{c|}{\textbf{A100 40G}} &  \multicolumn{4}{c}{\textbf{A100 80G}} \\ 
            \cline{2-9}
            
            & \textbf{1} & \textbf{2} & \textbf{4} & \textbf{8} & \textbf{4} & \textbf{8} & \textbf{16} & \textbf{32} \\
            \midrule[0.5pt]
            \textbf{2.7B} & 128K & 512K & 2M & 4M & 4M & 8M+ & 8M+ & 8M+ \\
            \hline
            \textbf{8B} & - & - & - & 1M & 2M & 4M & 8M+ & 8M+ \\
            \hline
            \textbf{13B} & - & - & - & 256K & 512K & 3M & 4M & 8M+ \\
            \hline
            \textbf{30B} &- & - & - & - & - & 1M & 3M & 4M \\
            \hline
            \textbf{70B} & - & - & - & - & - & - & 1M & 4M \\
            \toprule[1pt]
        \end{tabular}
    }
\caption{Maximum context length supported for LLM training with our FPDT design. \textbf{+} denotes we only test to this length. \textbf{-} denotes that the model itself cannot fit into GPU memory.}
\vspace{-1em}
\label{tab:max_len}
\end{table*}

There are multiple difficulties in training long-context LLMs. For a given-sized LLM model, as we increase the sequence length during training, the memory needed for storing activations and intermediate buffers increases proportionally, 
while the model parameters and optimizer states remain at a certain point. As we identified in this paper, this increase in activation memory can lead to severe GPU memory pressure. Previous memory-efficient techniques, such as FlashAttention~\cite{dao2022flashattention, dao2023flashattention}, have been proposed to alleviate the memory burden of materializing the giant $QK^T$ matrix, reducing memory complexity from $O(N^2)$ to $O(N)$, though, it still has a non-trivial constant factor which can easily cause out-of-memory issues when the sequence length grows to millions of tokens.
Methods like Megatron-SP~\cite{korthikanti2023reducing} and DeepSpeed Ulysses~\cite{jacobs2023deepspeed} have been proposed to leverage distributed GPU clusters. Megatron-SP adopts tensor parallelism to distribute the computation and memory of long sequences. In contrast, DeepSpeed Ulysses leverages the multi-head attention feature in current LLM models, using efficient all-to-all communication to distribute context head-wise, thereby easing memory pressure.

These sequence parallel strategies, despite being proven the feasibility of training LLMs with long contexts, require a substantial number of GPUs. For instance, training a 7B parameter model with a context window of 256K tokens using Megatron-SP results requires more than 32 A100 80G GPUs. Similarly, training a small 1.2B parameter GPT model with a 1M token context length using DeepSpeed Ulysses requires 64 A100 GPUs. These costly hardware pose severe usability challenges for companies and individual researchers with limited resources. Although with existing techniques such as activation checkpoint, the above solutions can extend further to longer sequence length, we identify that none of them can meet the expectation.

In this work, we aim to address these challenges by proposing a new \textbf{Fully Pipelined Distributed Transformer (FPDT)} for training long-context LLMs. Our method leverages the multiple memory hierarchies in modern GPU clusters, enhancing their hardware efficiency and cost-effectiveness while reaching extremely high MFU.

Our contributions are as follows:
\begin{itemize}
    \item We provide an end-to-end analysis of the memory footprint of LLM training, identifying memory spikes concerning commonly used Transformer architectures, and targeting on reducing redundant intermediate buffers in forward and backward passes respectively.
    \item We design a fully pipelined distributed transformer based on DeepSpeed Ulysses, tailored for \textbf{LLMs with sequence lengths of millions of tokens}, leveraging both GPU and host CPU memory as well as prefetching, to achieve a near-zero overhead training flow.
    \item We significantly reduce the GPU memory footprint of activations during the training of LLM, while leveraging a dedicated double buffer design to overlap almost all prefetching with computation. 
    \item As shown in Table~\ref{tab:max_len}, our proposed method supports training 8B LLMs on over 2M sequence with only 4 GPUs, or 70B models on 4M sequence with 32 GPUs, which can be up to 16x longer than existing solutions, while reaching over 55\% of MFU.
    \item Our solution works orthogonally and composably with DeepSpeed ZeRO family~\cite{rajbhandari2020zero} and PyTorch FSDP~\cite{zhao2023pytorch} of memory optimizations, and can be applied to Transformer-based models of any size, such as GPT, Llama, etc.
\end{itemize}

\section{Related Work}

\subsection{Memory-efficient Transformer}
The substantial memory demands of Transformers have spurred extensive research into memory-efficient attention mechanisms to facilitate their application to longer sequences. FlashAttention~\cite{dao2022flashattention} utilizes an online softmax computation technique~\cite{milakov2018online} and reduces memory requirements of self-attention from \( O(n^2) \) to \( O(n) \)~\cite{rabe2021self} while preserving the accuracy. Other notable strategies include low-rank approximations~\cite{wang2020linformer, katharopoulos2020transformers}, kernel-based methods~\cite{kitaev2020reformer, xiong2021nystromformer, lu2021soft}, and sparse attention mechanisms~\cite{child2019generating}, which approximate or selectively compute attention to minimize memory consumption. 
Furthermore, techniques that combine local and global contexts~\cite{ainslie2020etc, liu2021swin, beltagy2020longformer, zaheer2020big} enable superior performance on tasks involving long sequences or large-scale inputs while maintaining computational efficiency.
Our work is inspired by FlashAttention and builds upon it by introducing hierarchical blockwise computation, leveraging the memory hierarchy in modern systems, which significantly enhances scalability and reduces memory requirements. 

\subsection{Long context training}

Recent advancements in Transformer architectures have significantly enhanced their capability to process long sequences, which is crucial for tasks that require extensive contextual understanding. This section reviews pivotal contributions in this domain, each addressing the inherent memory limitations of standard Transformer models, while also pointing out some of their practical challenges.

Megatron-SP\cite{korthikanti2023reducing} adopts a sequence parallelism technique which is tightly integrated with its tensor parallelism. In this approach, sequences are partitioned along the sequence dimension, and all-gather and reduce-scatter collectives are employed to aggregate the QKV (query, key, value) projections for attention computation. The communication complexity analysis indicates that, in contrast to our approach, the communication volume in Megatron-SP's sequence parallelism increases linearly with the sequence length regardless of the number of compute devices.

The Blockwise Parallel Transformer (BPT) \cite{liu2024blockwise} employs a blockwise computation strategy for both self-attention and feedforward layers, optimizing memory usage and allowing the processing of sequences much longer than traditional Transformers. However, despite its efficiency, BPT requires careful tuning of block sizes and memory management to avoid diminishing returns on performance when scaling to extremely long sequences.

Ring Attention \cite{liu2023ring} enhances Transformer's scalability by distributing long sequences across multiple devices. This innovative approach overlaps the communication of key-value pairs with the computation of blockwise attention, effectively increasing the feasible sequence length proportionally to the number of available devices. However, reliance on device count for scaling and multi-step communications introduces potential issues in environments with sub-optimal hardware regards network interconnects, where performance can be unpredictably affected by network latency and bandwidth constraints.

DeepSpeed Ulysses \cite{jacobs2023deepspeed} tackles the challenges of sequence parallelism by partitioning input data along the sequence dimension and utilizing an efficient all-to-all collective communication strategy for attention computations. Although this method maintains a constant communication volume regardless of the increase in sequence lengths and device counts, achieving significant speedups and scalability, it may still encounter practical hurdles in deployment related to large-scale clusters and the optimization of communication patterns across diverse computing environments.

Each of these methods uniquely contributes to the field of long-sequence Transformer training, offering solutions focused on memory efficiency, communication overhead, and computational speed. However, the common requirement for substantial GPU resources in all these approaches underscores a critical barrier to broader adoption in scientific research and deployment. This challenge highlights the need for future research to develop more resource-efficient methods that can deliver similar benefits with a modest hardware budget.

Besides above mentioned large-scale training techniques, several recent works focus on long-sequence LLM training on small-scale GPU clusters. MsT~\cite{luo2024mini} adopts a chunking mechanism on MLP and loss computation, which showed benefits in reducing GPU memory consumption. However, as we will mention later, attention computation can incur the most significant memory spikes during training, which remains unsolved in their method. MEMO~\cite{zhao2024efficiently}, on the other hand, targets reducing memory in the attention part by using host memory for offloading. Their method requires integer programming solvers, and since it is built up on the tensor parallel scheme, failing to achieve optimal hardware efficiency in long-sequence training scenarios. 

\section{Preliminary}

\subsection{GPU memory requirements in distributed Transformer}

\begin{table}[H]
    \centering
    \vspace{-1em}
    \caption{Memory footprint at each step in a Transformer block}
    \label{tab:memory_transformer}
    \begin{tabular}{*{7}{c}}  
        \toprule
         & 
        \rotatebox{90}{Hidden} & 
        \rotatebox{90}{QKV proj.} & 
        \rotatebox{90}{All2all} & 
        \rotatebox{90}{Attention} & 
        \rotatebox{90}{FFN} &
        \rotatebox{90}{Other ops.}  \\[1ex]  
        
        \midrule
        \multicolumn{7}{c}{Activations} \\
        \midrule
        Forward & $Nd$ & $3Nd$ & \multirow{2}{*}{$4Nd$}& $4Nd$ & $4Nd$  & \multirow{2}{*}{$3Nd$} \\
        Backward & $2Nd$ & $6Nd$ & & $8Nd$& $8Nd$&  \\
        \bottomrule
    \end{tabular}
    \vspace{-1em}
\end{table}
As Transformer becomes the de facto most powerful and ubiquitous architecture in today's generative models, its overall pipeline also becomes similar. Table~\ref{tab:memory_transformer} shows the required GPU memory at each step in the forward and backward of a Transformer block using DeepSpeed Ulysses.

Noticeable in this table, is that to get query, key, and value, the memory footprint is directly increased by three times, which solely can potentially lead to an out-of-memory issue when the sequence itself is too long to fit in the GPU memory.
Also, as most Alltoall implementations do not support in-place operation, we need to create a receive buffer in addition to the input buffer. And if asynchronous communication is enabled, we need to prepare the receive buffer for query, key, and value at one time, leading to a $6Nd$ memory footprint, where $d$ denotes the hidden dimension. 

FlashAttention is introduced to reduce the memory consumption to $O(N)$, however, in practice, it can also incur a huge memory footprint. For example, in backward, Flash-Attention requires the following inputs: $q$, $k$, $v$, $o_f$, $o_g$, $dq$, $dk$, $dv$, where $o_f$ denotes the attention output of the forward pass, and $o_g$ denotes the gradient of output in the backward pass. As these tensors need to be present in GPU memory at one time, it directly leads to a $8Nd$ memory footprint.
We identify that none of the existing solutions exclusively target these memory burdens, which, however, becomes salient in long-context LLM training.

\subsection{Combining DeepSpeed sequence parallel and ZeRO}
\label{sec:zeros_sp}
Among sequence parallel strategies, DeepSpeed-Ulysses excels with its highly efficient communication pattern and is complementary to the most advanced model-based training schemes such as DeepSpeed ZeRO. We first recap the key feature of DeepSpeed Ulysses, and how it can work with ZeRO-3. 

\begin{figure}[H]
    \centering
    \includegraphics[width=0.46\textwidth]{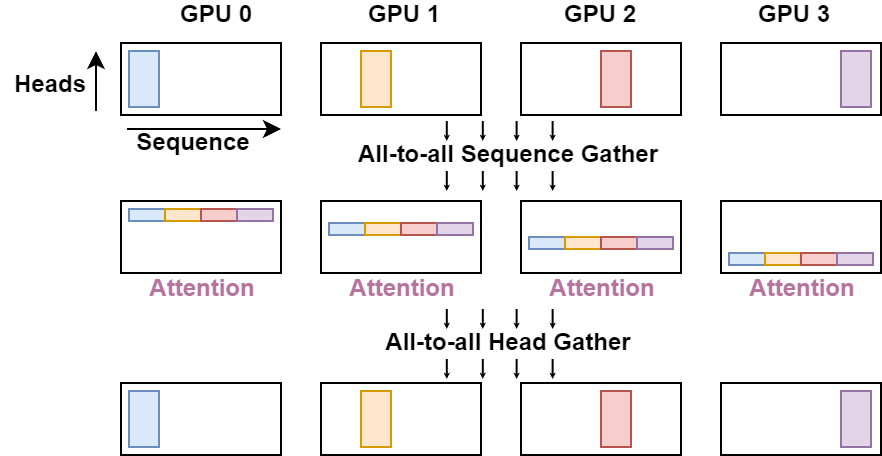}
    \caption{DeepSpeed Ulysses with distributed attention.}
    \label{fig:zero3}
    \vspace{-1em}
\end{figure}

\begin{figure}[H]
\vspace{-1em}
    \centering
    \includegraphics[width=0.46\textwidth]{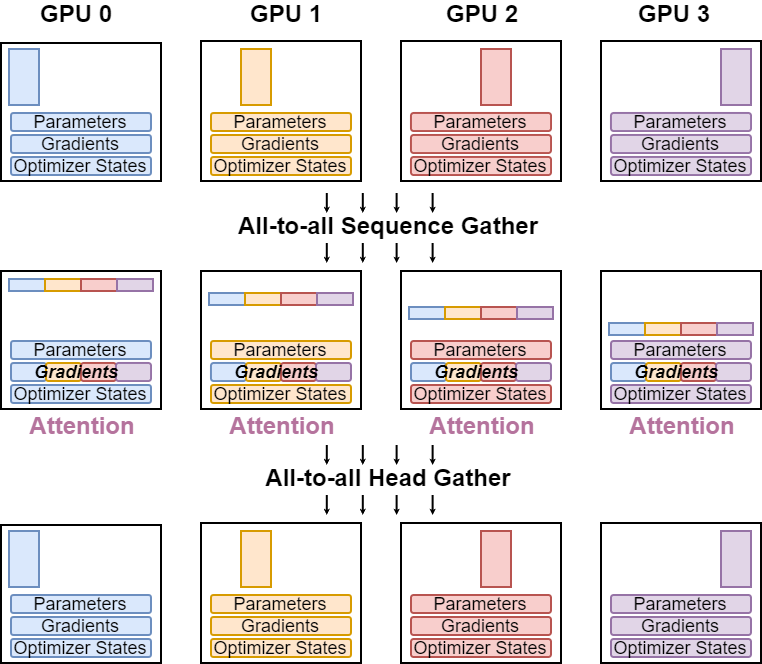}
    \caption{DeepSpeed Ulysses works orthogonally with ZeRO-3.}
    \label{fig:ulysses_zero3}
\vspace{-1em}
\end{figure}

Figure~\ref{fig:zero3} (a) shows the communication pattern of DeepSpeed Ulysses sequence parallelism. Each GPU initially has a piece of the entire sequence with full heads, when performing attention, sequence pieces are gathered to each GPU, while heads are scattered to different GPUs. ZeRO3 partitions all parameters, gradients, and optimizer states along a data-parallel GPU group. As sequences consist of tokens, they can be easily treated as batches of data, where we can directly apply the data parallelism. As shown in Fig~\ref{fig:ulysses_zero3}, in this combined training scheme, sequence parallel is used to reduce activation memory, and ZeRO-3 reduces the model-related memory.

\section{Design of Fully Pipelined Distributed Transformer}
In this section, we start by introducing how to design a seamless and efficient pipelined distributed attention leveraging both host and GPU memory. Since different hardware hierarchies are involved in the design, e.g. Tensor cores, NVLINK, PCIe, etc, each with different data throughput and latency, they are required to coordinate carefully.

\subsection{Pipelining and scheduling}

As QKV projection, Alltoall communication, attention, and FFN will create multiple intermediate buffers, leading to severe memory spikes, especially during the backward pass, to make the sequence computation in the Transformer block fully pipelined and memory efficient, our chunk and offloading design will start with the initial input tensor (i.e. hidden state). Firstly, we use the following notations throughout the paper for ease of explanation. For operations other than the distributed attention, each GPU holds and processes a $[b, s_{local}, h_{global}, d]$ tensor, where $s_{local}$ denotes the local sequence length, $h_{global}$
denotes the total number of heads. For the distributed attention, each GPU will process the entire sequence, but with specific heads, which is a $[b, s_{global}, h_{local}, d]$ tensor.

\begin{figure}[H]
    \centering
    \vspace{-1em}
    \includegraphics[width=0.47\textwidth]{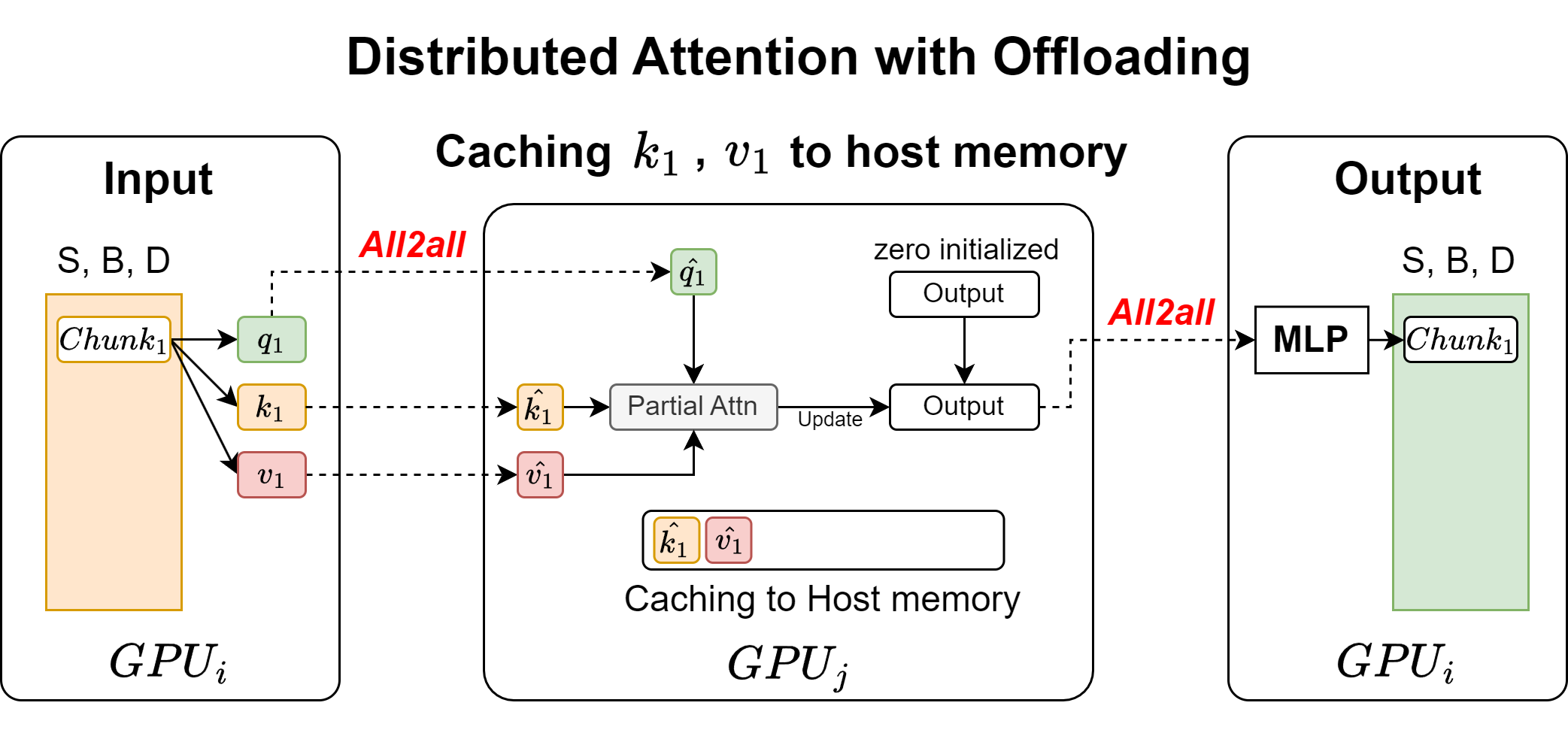}
    \vspace{-1em}
    \caption{The design of distributed attention with offloading}
    \label{fig:pipele_case0}
    \vspace{-1em}
\end{figure}

For the first QKV projection, since tokens are processed elementwise, we directly slice the local sequence tensor $[b, s_{local}, h_{global}, d]$ into $u$ chunks, each as a $[b, \frac{s_{local}}{u}, h_{global}, d]$ tensor. We denote these chunks as $T_i$, where $i \in \{0, 1, \ldots, u-1\}$. As shown in Figure~\ref{fig:pipele_case0}, $T_i$ is projected to query $q_i$, key $k_i$, and value $v_i$. Then, we perform the Alltoall communication among the sequence parallel group. Recall that in table~\ref{tab:memory_transformer}, we mentioned that since Alltoall communication cannot be performed in-place, additional receive buffers need to be pre-allocated. In our chunk design, as we largely reduced the sequence length by a factor of $u$, even though the receiving buffers for $\hat{q_i}, \hat{k_i},\hat{v_i}$ are still required, they become trivial compared to the original sequence tensor with the full context.

After using Alltoall to scatter heads and gather sequence, each chunk $\hat{q_i}, \hat{k_i},\hat{v_i}$ is a $[b, \frac{s_{global}}{u}, h_{local}, d]$ tensor. Note that, for generative LLMs, due to the casual mask, after the computation, $\hat{q_i}$ will not be used in the forward pass anymore, while $\hat{k_i},\hat{v_i}$ need to participate in the following $\hat{q_j}$ computation, where $i<j$. 
Therefore, we cache $\hat{k_i},\hat{v_i}$ to the host memory. 
Specifically, for $\hat{q_0}, \hat{k_0},\hat{v_0}$, we can directly get the final output of chunk $T_0$, as $\hat{q_0}$ will not attend to the remaining sequence. For chunk $T_i, i\geq 0$, processing each chunk only gives intermediate results, which will be rescaled in the next chunk computation. As online attention is widely used, we adopt a similar strategy when scheduling the attention computation.

\begin{figure}[H]
    \centering
    \includegraphics[width=0.47\textwidth]{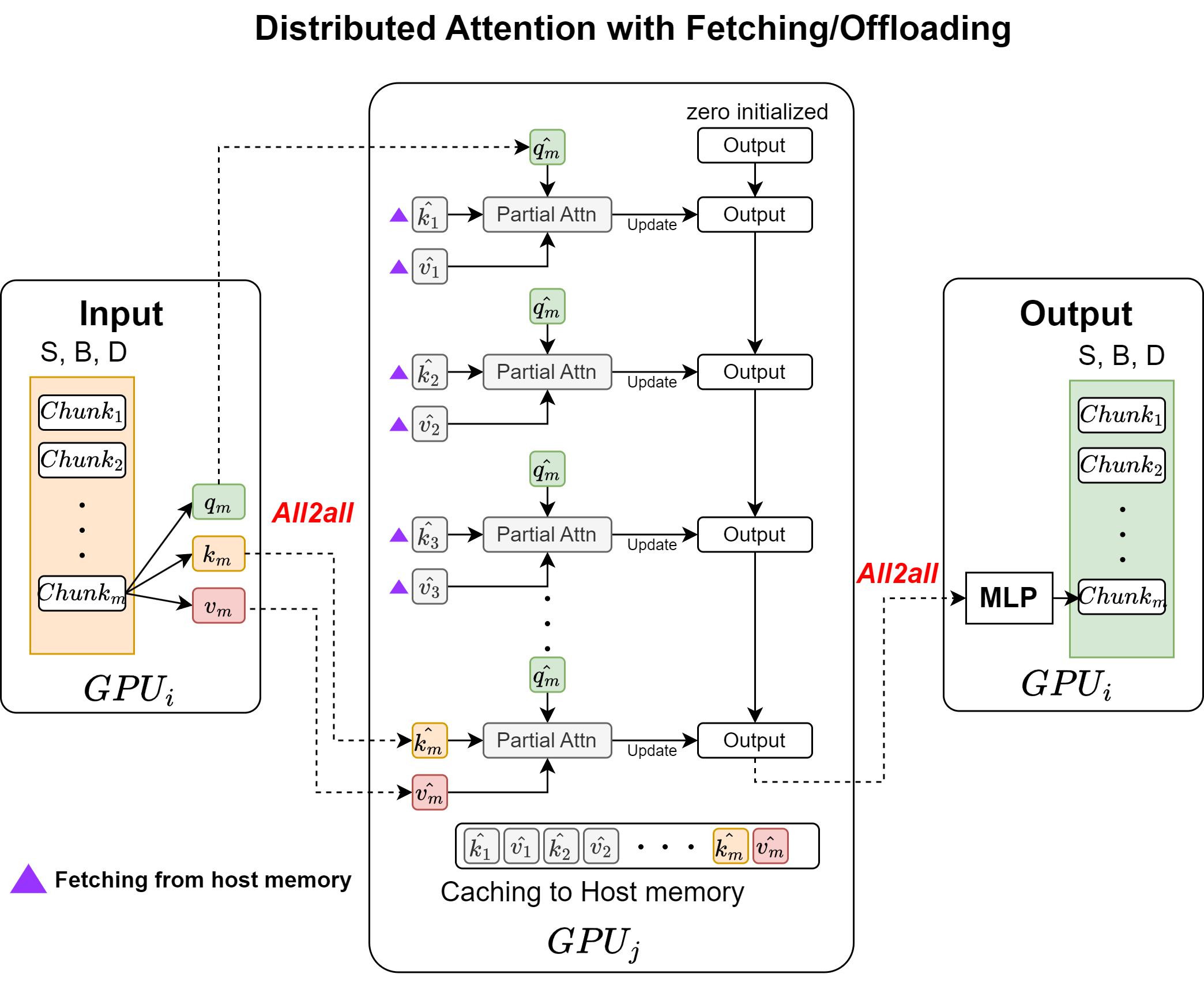}
    \vspace{-1em}
    \caption{The design of distributed attention with fetching and offloading. We follow the online attention policy to update attention output.}
    \label{fig:pipele_case2}
\vspace{-1em}
\end{figure}
 Figure~\ref{fig:pipele_case2} gives an example of how to perform the computation of chunk $T_m$. After the Alltoall operation, $GPU_j$ receives $\hat{q_m}$, $\hat{k_m}$, and $\hat{v_m}$. We then fetch the previous sequence chunk by chunk from the host memory to $GPU_j$, and perform online attention with the current $\hat{q_m}$, and update the output chunk accordingly. Note that, in a strict manner, at any given time, only one set of chunks $\hat{k_i}, \hat{v_i}$ is placed on GPU's HBM, reducing the memory footprint to $\frac{1}{u}$ compared to the non-offloading version.

 Unlike inference, where activations and intermediate results can be freed as soon as the forward pass of the current Transformer block finishes, training requires saving necessary intermediate outputs as it is reused in the backward pass. 
 Therefore, we offload $\hat{q_i}, \hat{k_i}, \hat{v_i}$ to the host memory once they are done for forward computation. 

\begin{figure}[H]
    \centering
    \includegraphics[width=0.47\textwidth]{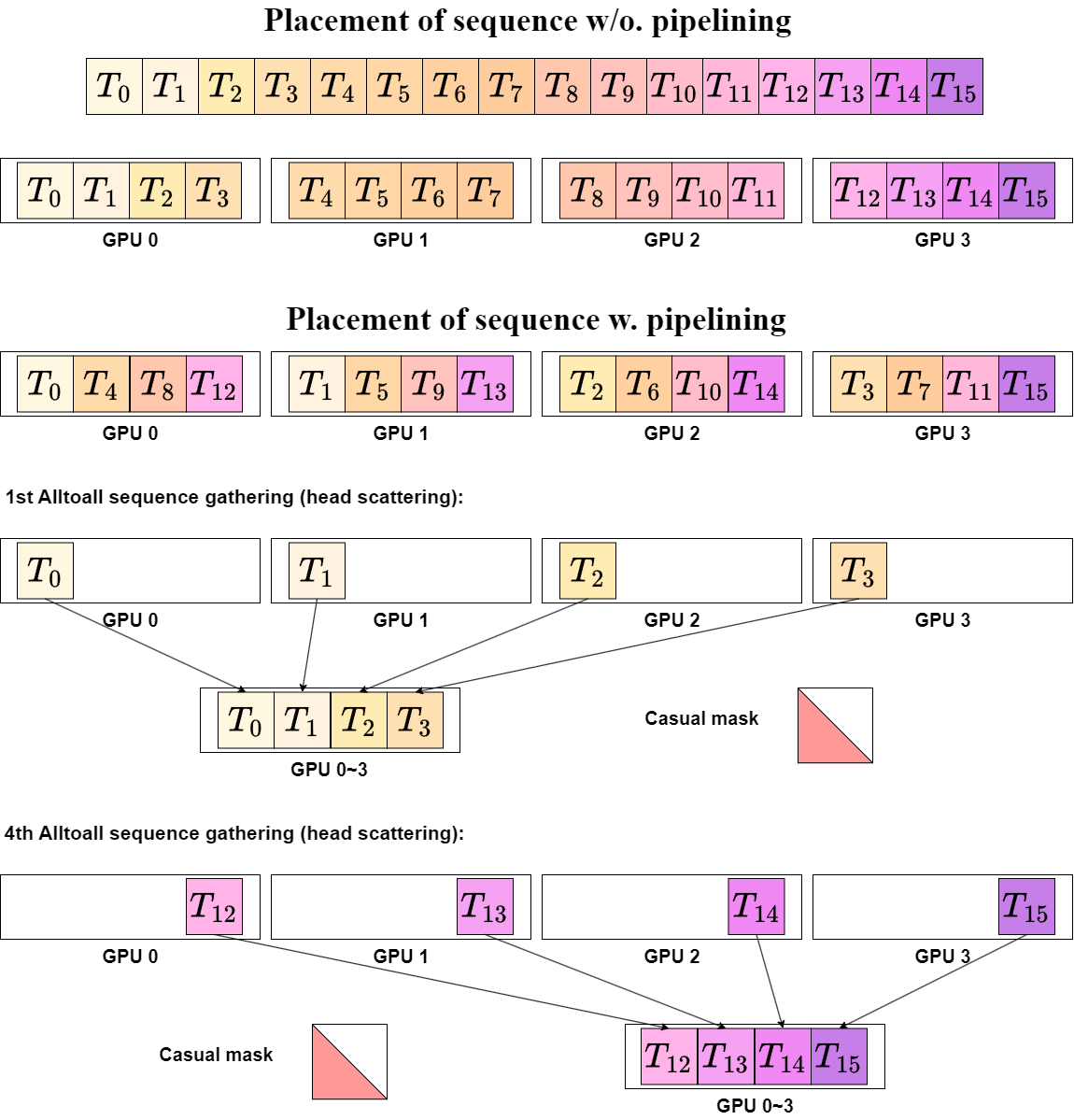}
    \caption{Rank ordinal scattering of sequence chunks. The diagonal casual mask remains valid after each chunk Alltoall operation. NVLINK is also load-balanced in this data layout.}
    \vspace{-1em}
    \label{fig:seq_shuffle}
\end{figure}

Before diving into more advanced designs, we need to shuffle the sequence accordingly due to the chunked Alltoall sequence gathering. In the original sequence parallel scheme, the input sequence is sliced and dispatched to each GPU based on their rank. As shown in Figure~\ref{fig:seq_shuffle}, if we directly perform the Alltoall operation on the $i_{th}$ chunk on each GPU, then the sequence we gather does not conform to the casual mask.
\begin{figure*}[htbp]
    \centering
    \includegraphics[width=0.98\textwidth, page=1]{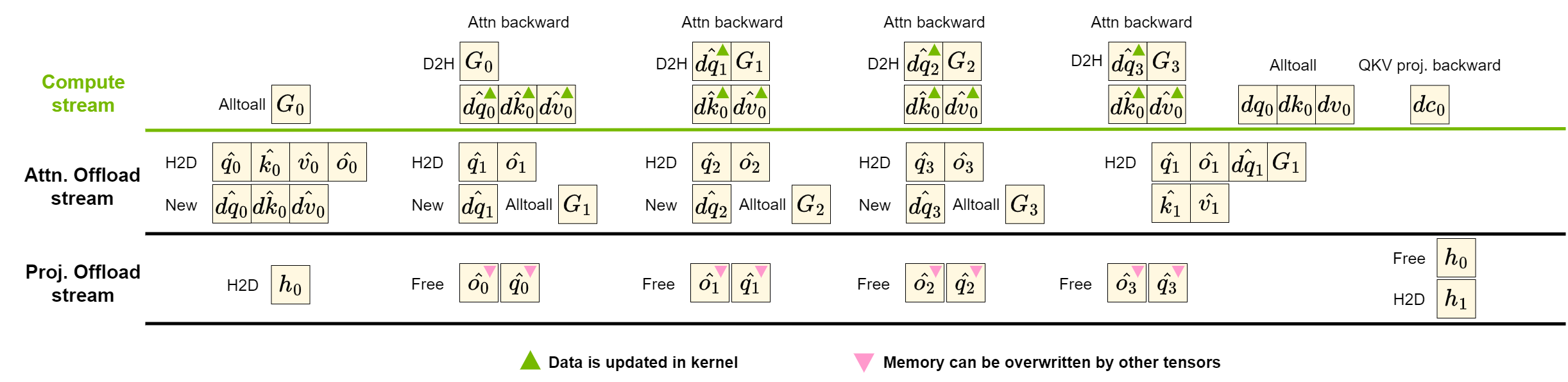}
    \caption{Double buffer leverages multiple CUDA streams in the backward pass of a Transformer block. We overlap most offloading operations with the attention gradients computation. We only show attention-related operations for clarity.}
    \label{fig:bw_db}
    \vspace{-1em}
\end{figure*}

For example, when gathering the second chunk, each GPU will get a sequence consisting of chunk $T_1, T_5, T_9, T_{13}$ with rank-specific heads. However, since each GPU has already received $T_0, T_4, T_8, T_{12}$ in the previous chunk computation, the casual mask for $\hat{q_1}, \hat{k_0}$ needs to be dedicatedly designed, as $T_5$ can only attend to tokens in $T_0$ and $T_4$, etc. Another workaround would be to only let $i_{th}$ GPU scatter its sequence to all GPUs, this does not require reordering the sequence, but will result in NVLINK load imbalance.
During training, since each GPU computes the loss of the sequence it holds, we also reorder the labels accordingly, so that the loss still matches. Note that we shuffle the input token ids and labels in the data loader; thus, there is no overhead in this reordering of sequence.

We also emphasize that, unlike Ring Attention, in our design, each GPU always processes the same piece of sequence at any given time, with the only difference in rank-specific heads. Therefore, GPUs are always load-balanced when computing attention. This uniform workflow also reduces the synchronization and barrier required to coordinate the sequence of the parallel group.

\subsection{Double buffering}
\label{sec:double_buffer}
Though the idea of using host memory to hold unused sequences is intuitive, the unmatched hardware transfer bandwidth poses a significant challenge in fully exploiting computing power.
For a typical HPC node, GPUs are connected through high-bandwidth NVLink, which can reach more than 100 GB/s of peer-to-peer bandwidth. However, the common PCIe Gen-4 link with 16 lanes only provides 32 GB/s of unidirectional bandwidth, which also requires the host memory and GPU to be in the same NUMA domain. 

\begin{figure}[H]
    \centering
    \begin{subfigure}[t]{0.28\textwidth}
        \centering
        \includegraphics[width=\textwidth]{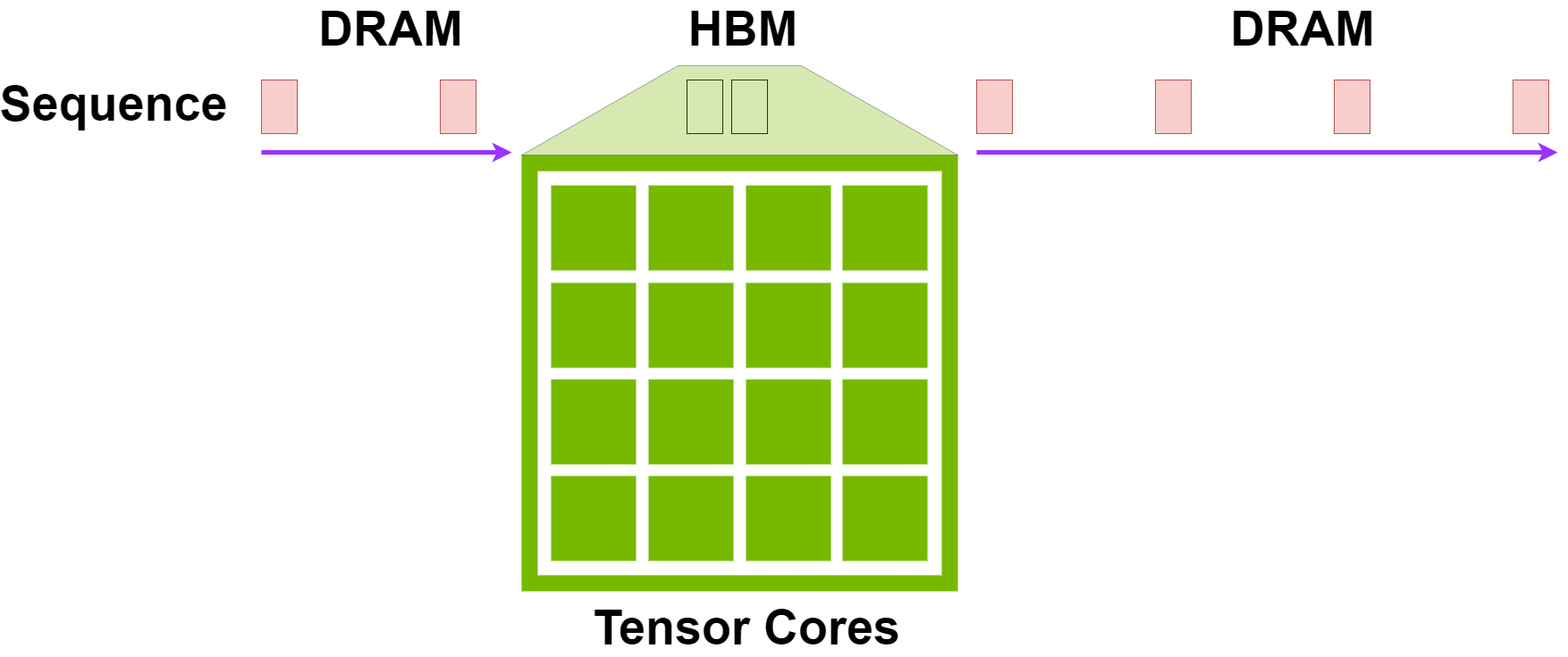}
        \label{fig:db_0}
    \end{subfigure}%
    \hfill
    \begin{subfigure}[t]{0.17\textwidth}
        \centering
        \includegraphics[width=\textwidth]{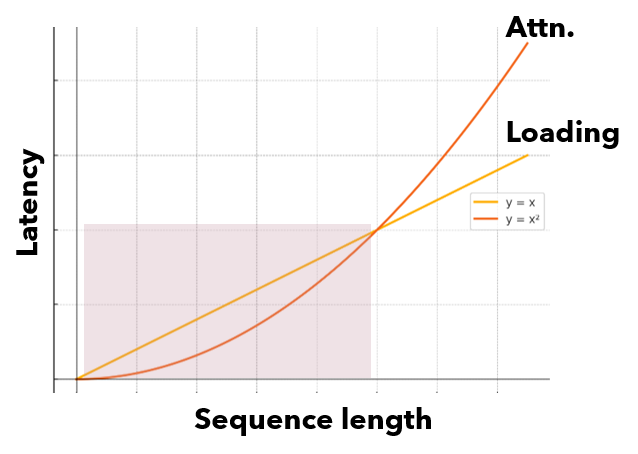}
        \label{fig:db_0_g}
    \end{subfigure}
    \vspace{-1em}
    \caption{Short sequence chunk leads to GPU starving.}
    \label{fig:starving}
    \vspace{-1em}
\end{figure}

Figure~\ref{fig:starving} shows a GPU starving case where the sequence chunk is too short such that the latency in the attention computation is less than the data fetch latency. In this situation, the training will be bound by the PCIe bandwidth, leading to low Model FLOPs Utilization (MFU). 

\vspace{-0.5em}
\begin{figure}[H]
    \centering
    \begin{subfigure}[t]{0.28\textwidth}
        \centering
        \includegraphics[width=\textwidth]{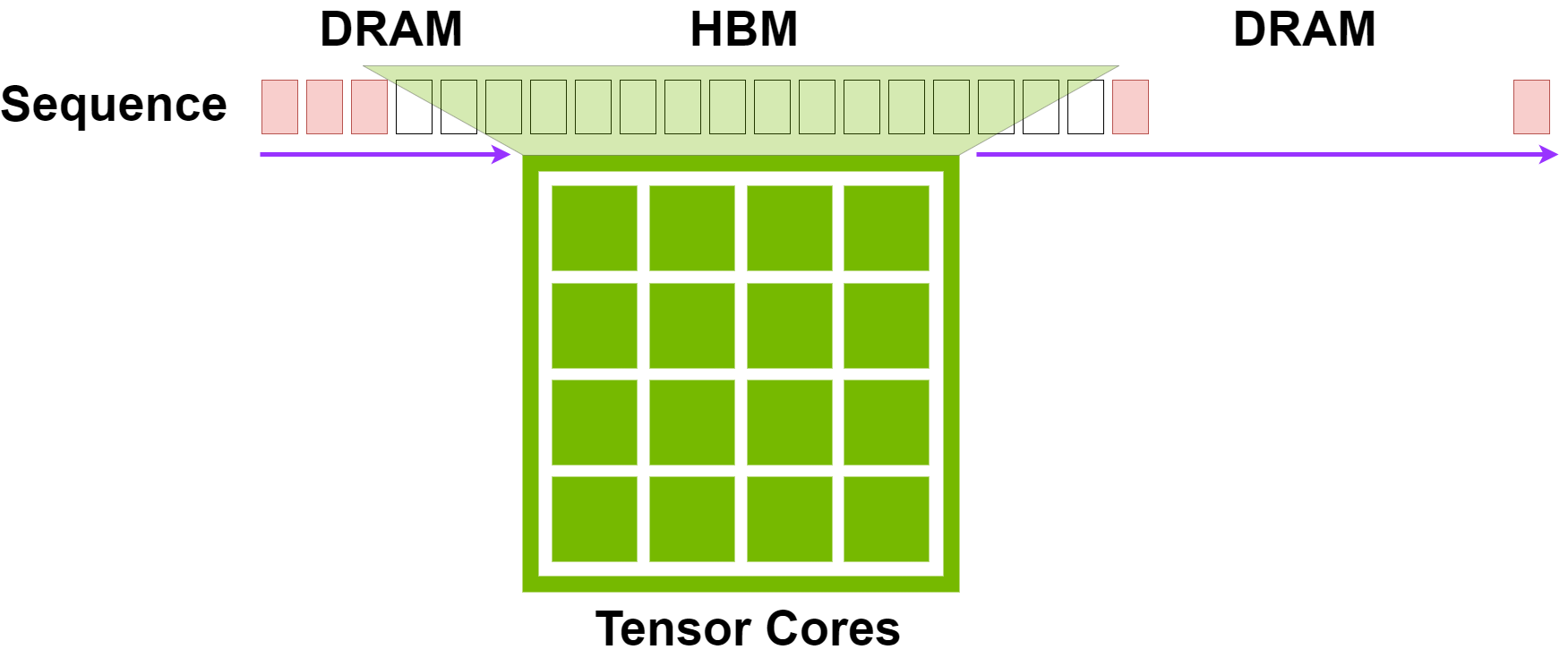}
        \label{fig:db_1}
    \end{subfigure}%
    \hfill
    \begin{subfigure}[t]{0.17\textwidth}
        \centering
        \includegraphics[width=\textwidth]{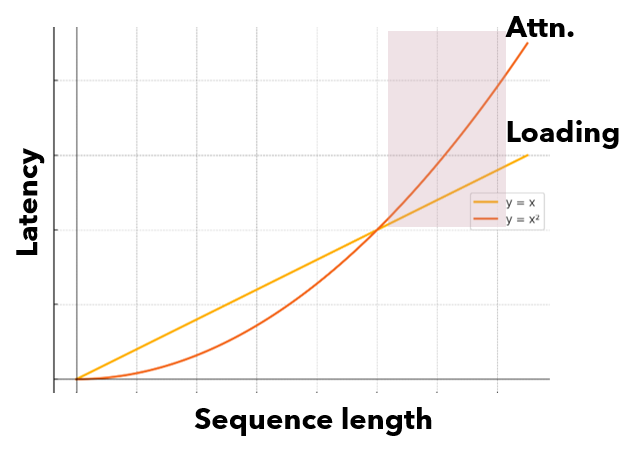}
        \label{fig:db_1_g}
    \end{subfigure}
    \vspace{-1em}
    \caption{Long sequence chunk leads to HBM wasting.}
    \label{fig:hbm_wasting}
\end{figure}
    \vspace{-1em}

Figure~\ref{fig:hbm_wasting} shows an opposite case where each sequence chunk is too long. Because the tensor we passed to a kernel (i.e. attention forward/backward kernels) needs to reside on HBM as a whole, it will take too much GPU memory, which becomes unnecessary, as tokens can be fetched from host memory as soon as previous ones are computed.

Furthermore, the mismatch in GPU computing throughput and PCIe link bandwidth also poses a severe difficulty in efficiently incorporating offloading into the regular computation pipeline. Figure~\ref{fig:avg_time} shows the latency of different operations performed on our GPU node (see \ref{kn:machine} for details). Note that here we compare the actual tensor operations in a transformer block, i.e, Alltoall on $[b, \frac{s}{p}, h, d]$ tensor, attention on $[b, s, \frac{h}{p}, d]$ tensor, and fetching on $[3, b, s, \frac{h}{p}, d]$ tensor which represents the query, key, and value. As the complexity of attention computation increases quadratically, it easily dominates the overall latency if we let each sequence chunk be large enough (e.g. 512k), in which case the latency of communication and fetching becomes negligible. In practice, however, a large chunk size also increases the memory pressure, which is against our initial goal. Finding the sweet point between fully utilizing computing cores and hiding the latency of offloading/fetching is crucial for achieving the optimal training efficiency. As shown in Figure~\ref{fig:avg_time}, Alltoall is much faster since this is only the intra-node communication using NVLink. 

Thus, we care more about when the latency of attention computation overpasses that of host-to-device fetching. Unlike inter-GPU connections, data transfer among host memory and GPUs is not homogeneous. In our case, all GPUs will share the PCIe bandwidth, especially when host-to-device memory copy is issued simultaneously on each GPU. 

\begin{figure}[H]
    \centering
    \includegraphics[width=0.42\textwidth]{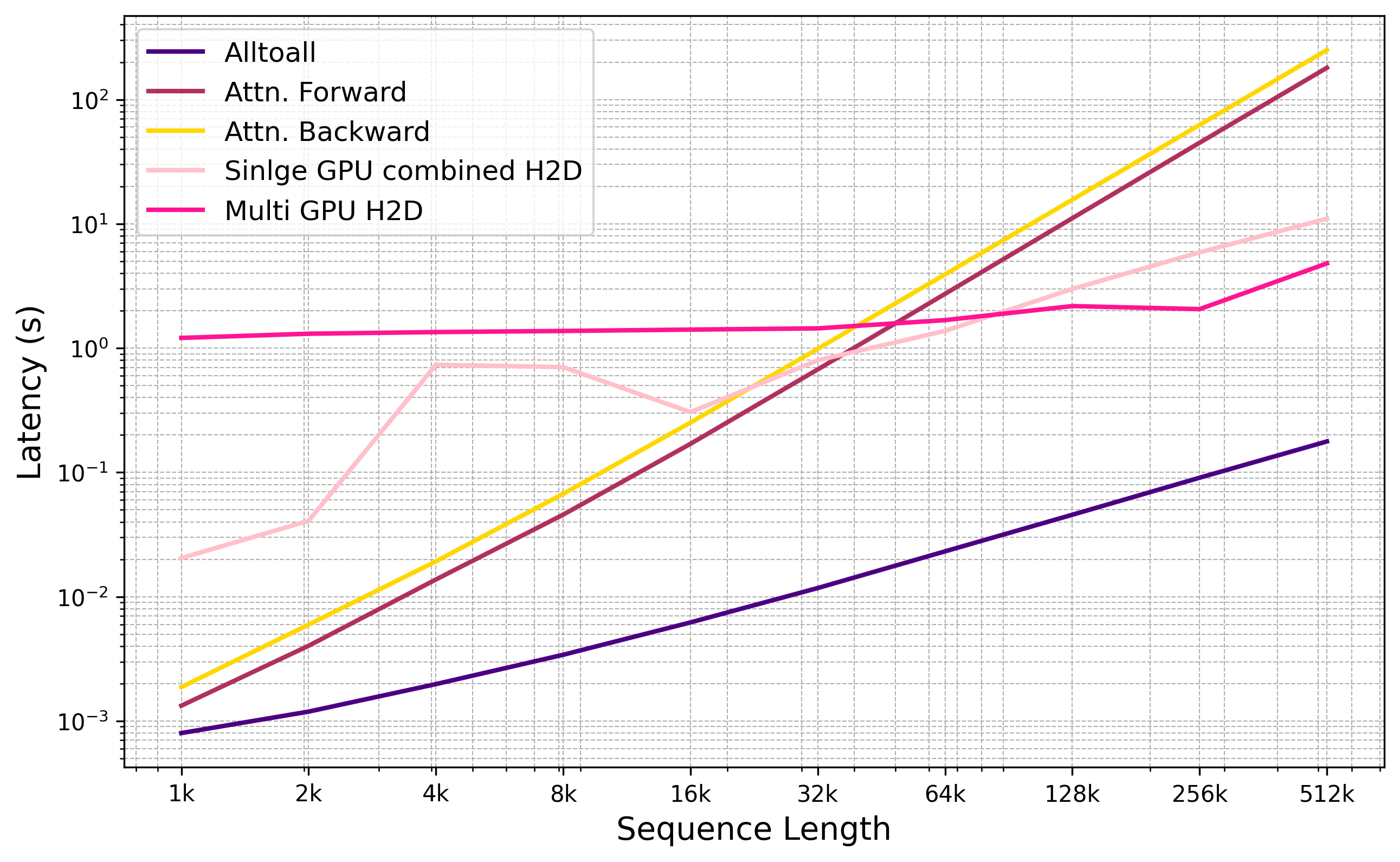}
    \caption{Average Time Spent in Alltoall, Attention Forward, Attention Backward, and two different host-to-device strategies.}
    \label{fig:avg_time}
    \vspace{-1em}
\end{figure}
Therefore, we investigated two different strategies for fetching data from host memory to the corresponding GPUs. The most intuitive way is to let each GPU issue its own $HtoD$ transferring, which can make use of all DMA engines available, however, this might lead to resource contention as each GPU will try to consume some PCIe bandwidth as well as PCIe lanes. The second strategy is to let one GPU in each node fetch all related tensors, and then use a scatter operation to send corresponding chunks to other devices. In our profiling, we found that though the multi-GPU $HtoD$ strategy performs worse at smaller data sizes, due to the overhead in lane contention, latencies of both methods are overpassed by attention computation at around 32k to 64k, and thus, their difference becomes negligible as the sequence length grows larger. 
Since fetching from one GPU and scattering to other GPUs require additional synchronization and barrier, we choose the first strategy where we allow every GPU to fetch its data without any additional synchronization.

\begin{figure*}[t]
    \centering
        \begin{subfigure}{0.32\textwidth}
            \centering
            \includegraphics[width=\linewidth]{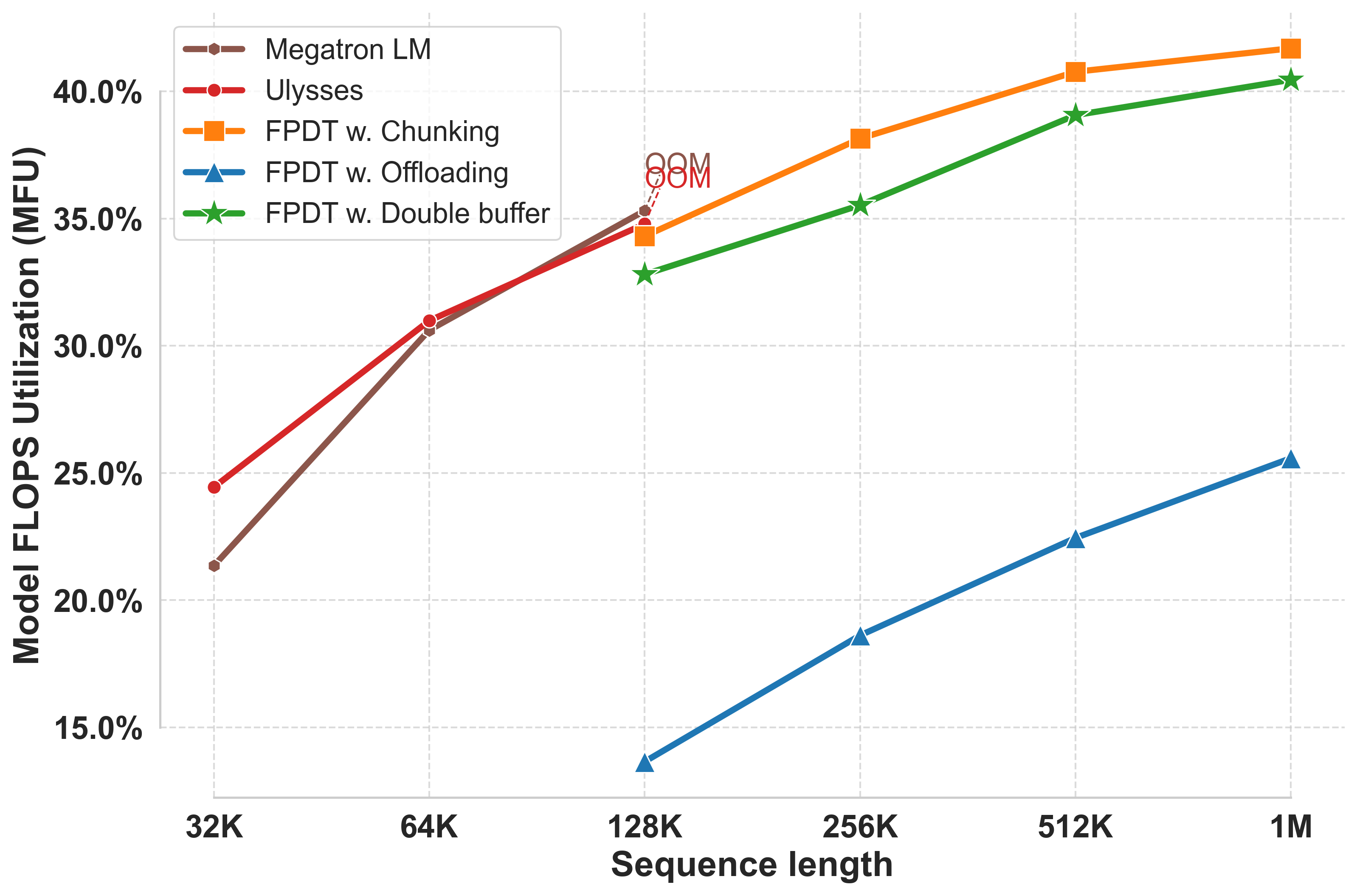}
            \caption{2.7B GPT, Num. of GPUs = 2}
        \end{subfigure} 
        \begin{subfigure}{0.32\textwidth}
            \centering
            \includegraphics[width=\linewidth]{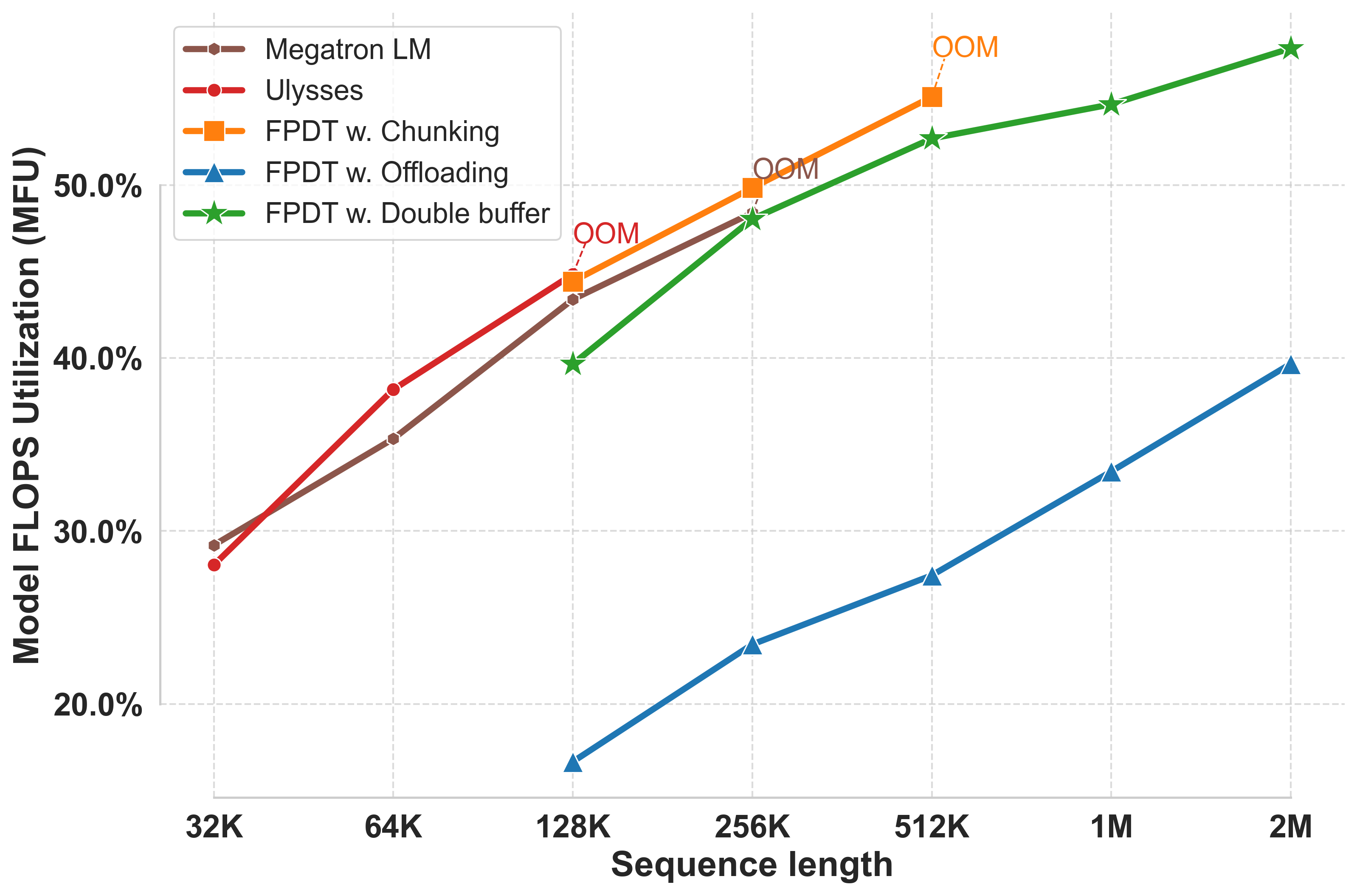}
            \caption{6.7B GPT, Num. of GPUs = 4}
        \end{subfigure} 
        \begin{subfigure}{0.32\textwidth}
            \centering
            \includegraphics[width=\linewidth]{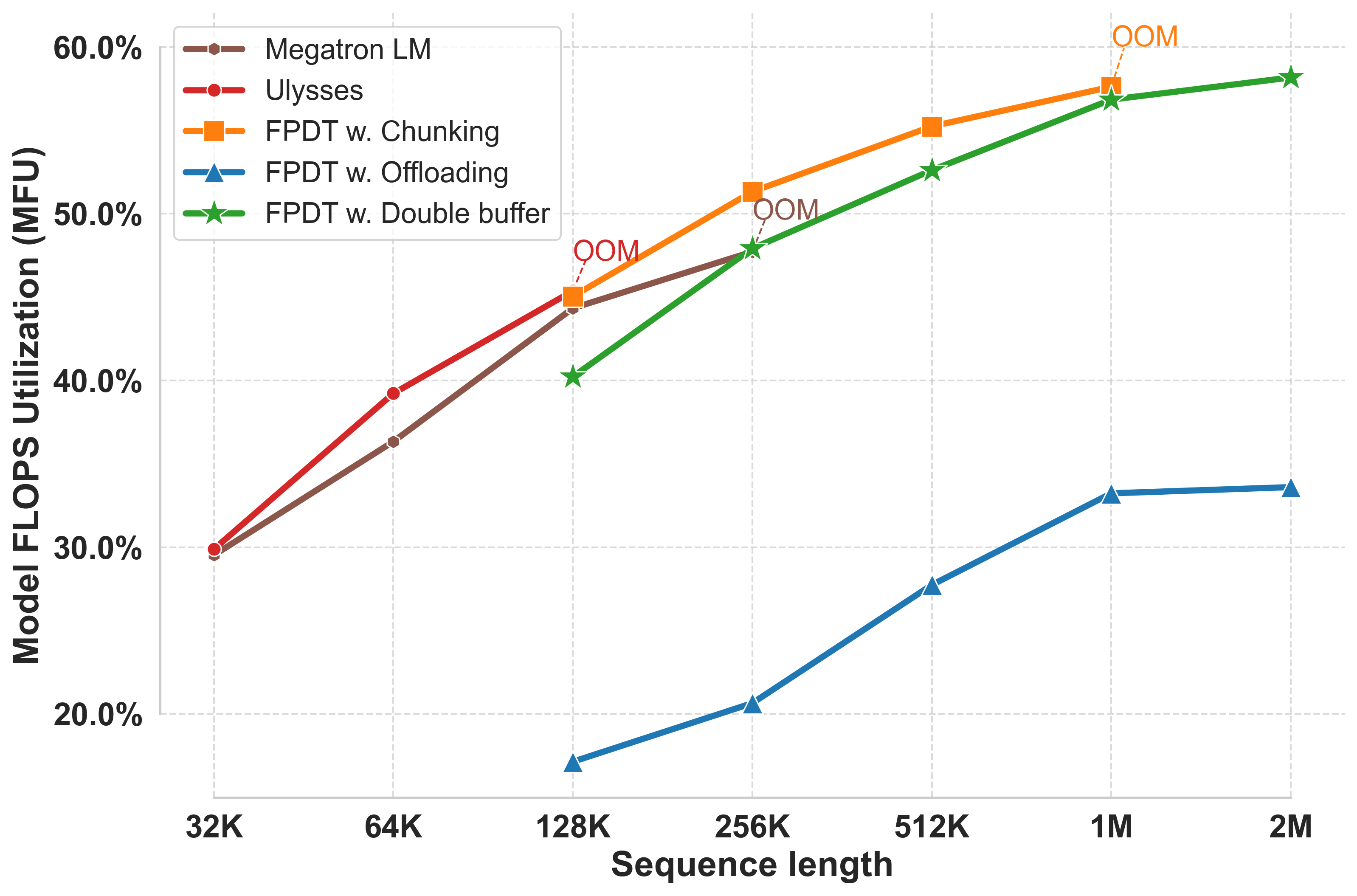}
            \caption{8B Llama 3, Num. of GPUs = 4}
        \end{subfigure}
        \begin{subfigure}{0.32\textwidth}
            \centering
            \includegraphics[width=\linewidth]{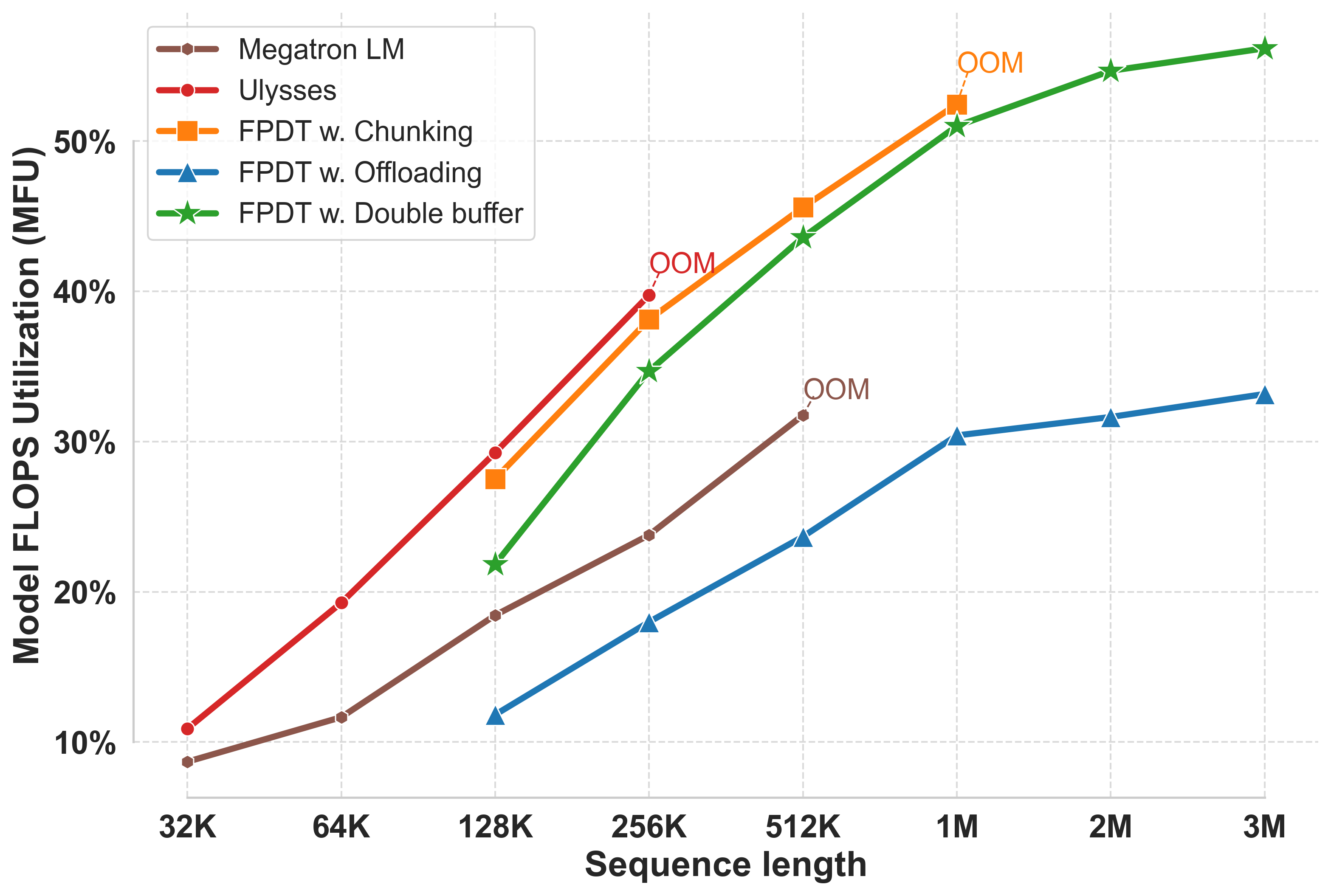}
            \caption{13B GPT, Num. of GPUs = 8}
        \end{subfigure} 
        \begin{subfigure}{0.32\textwidth}
            \centering
            \includegraphics[width=\linewidth]{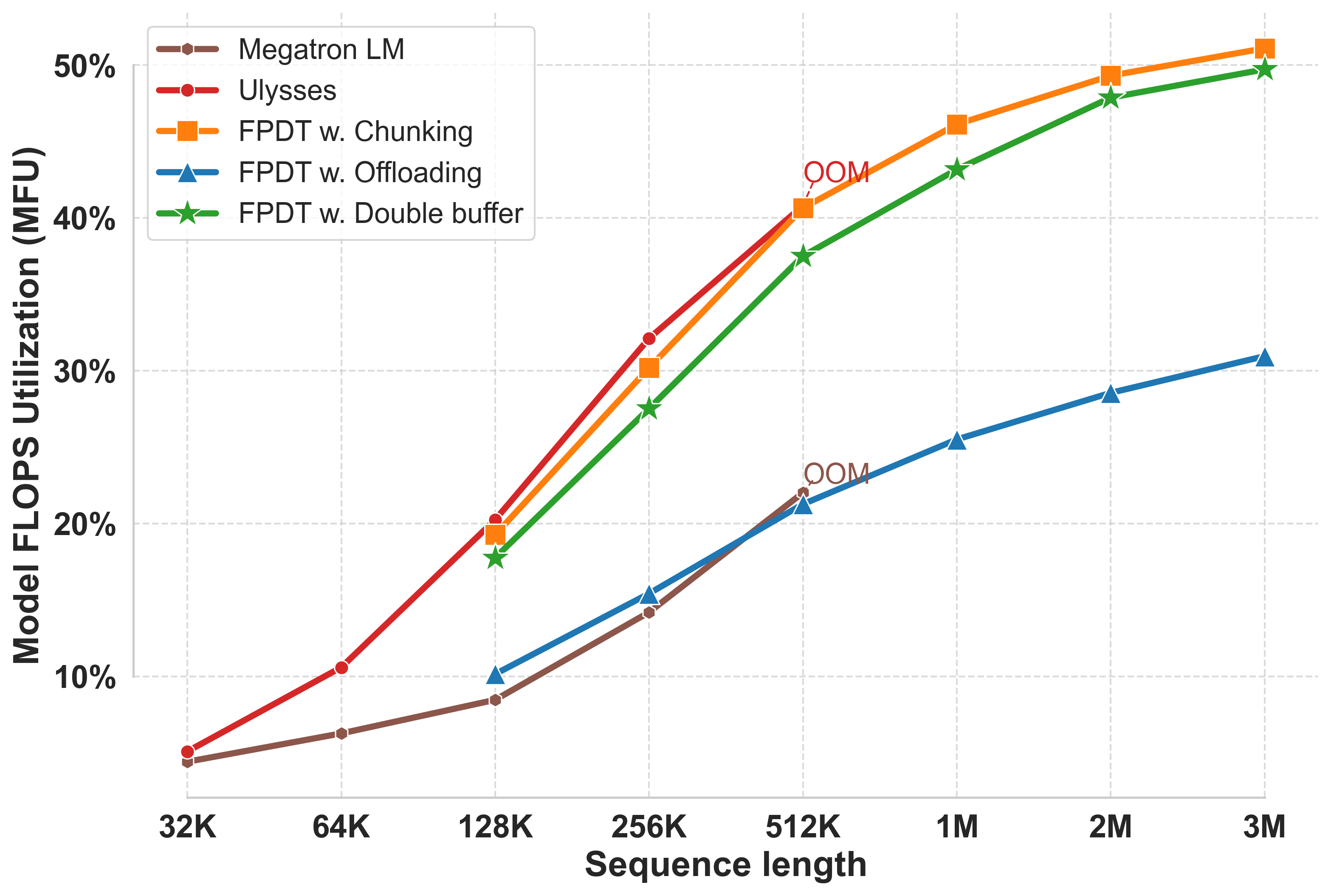}
            \caption{30B GPT, Num. of GPUs = 16}
        \end{subfigure} 
        \begin{subfigure}{0.32\textwidth}
            \centering
            \includegraphics[width=\linewidth]{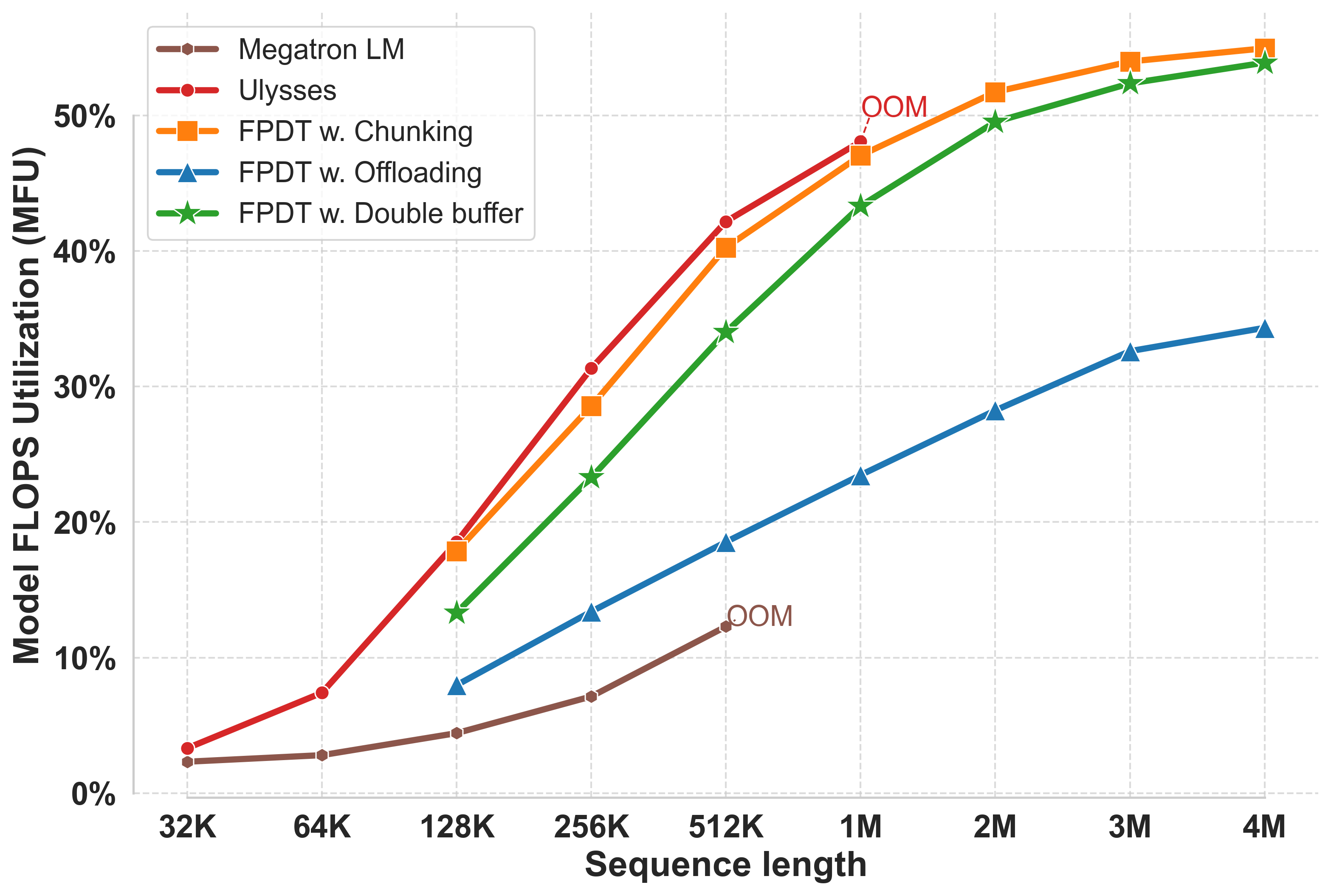}
            \caption{70B Llama 3, Num. of GPUs = 32}
        \end{subfigure} 
    \caption{Supported sequence lengths and corresponding Model FLOPs Utilization (MFU) using Megatron-SP, Ulysses, and our proposed FPDT. OOM denotes the point where increasing sequence length will cause memory issues. We show FPDT's performance when the sequence length is larger than 128K, as shorter sequences can be properly handled by existing strategies.}
    \vspace{-1em}
    \label{fig:mfu}
\end{figure*}

Figure~\ref{fig:bw_db} illustrates our double buffer workflow. Note that here we show the backward pass of our FPDT block, as the forward workflow is simpler and can also be reflected by examining the backward pass. We assume that there are 4 chunks in total. Since the global sequence chunk $\hat{q_i}$, $\hat{k_i}$, $\hat{v_i}$ have been cached during the forward, we then directly fetch them from the host memory without introducing additional Alltoall. We use a nested loop to compute and update the gradients of query $\hat{dq_i}$, key $\hat{dk_i}$, and value $\hat{dv_i}$. The outer loop is on key and value, while the inner one is on query. This design ensures that the attention computation in the inner loop only needs to cover the latency of fetching the next query, otherwise, it will need to cover both key and value prefetching. In the inner loop, since $\hat{k_0}$ and $\hat{v_0}$ are used by $\hat{q_0}$, $\hat{q_1}$, $\hat{q_2}$, $\hat{q_3}$, we therefore update $\hat{dk_0}$, $\hat{dv_0}$ each time. For $\hat{dq_0}$, we get its final result after the first inner loop, which is due to the autoregressive nature of LLM that $q_i$ never attends to $k_j$ if $i<j$. Similarly, we get the final results of $\hat{dk_0}$ and $\hat{dv_0}$ after the first outer loop. Then, we initiate Alltoall to scatter the global sequence chunk to its original GPU, where $dq_0$, $dk_0$, $dv_0$ are used to compute the gradient of the input hidden state $dc_0$. Note that here we also overlap Alltoall and projection backward with the prefetching of $\hat{q_1}$, $\hat{k_1}$, and $\hat{v_1}$, which will be used in the next outer loop. In our implementation, we deploy three CUDA streams, as the prefetching of the input hidden state $h_0$ will only be synced in the projection backward. For chunk buffers that will never be used after each inner loop, we mark them as free memory, which can be allocated to the following chunks.

\section{Evaluation}

\subsection{Experimental Setup}
\textbf{Models:} We conduct our main experiments using the GPT and Llama models, with model sizes ranging from 2.7B to 70B. By default, we enable activation checkpoint with CPU offloading. To fully exploit the potential of FPDT, we use DeepSpeed ZeRO-3 to partition the model parameters across the sequence parallel group (refer to~\ref{sec:zeros_sp}). We also set the batch size to 1, as this allows us to test the maximum sequence length we can reach.

\label{kn:machine}
\textbf{Experimental Environment:} We use multiple GPU nodes, each with four A100 80 GB, connected via 3rd-Gen NVLink. There are two CPU sockets. The PCIe between host and device has a theoretical unidirectional bandwidth of 32 GB/s. Each node is equipped with 1 TB of host memory. For the internode connection, we use NVIDIA 200 Gbps HDR InfiniBand.

\subsection{Overall Performance}

There are several widely used solutions for training long-context language models. Megatron-SP~\cite{korthikanti2023reducing} partitions sequence activations and leverages tensor parallel. DeepSpeed Ulysses~\cite{jacobs2023deepspeed} adopts a one-step Alltoall to gather tokens and scatter heads among all GPUs. Ring Attention leverages the online attention mechanism to distribute KV chunks and overlap the ring communication with local attention computation. However, the only end-to-end training framework with Ring Attention is Megatron-LM, which does not support activation checkpointing with CPU offloading. As we will also show in Table~\ref{tab:comprehensive_strategies}, this severely limits the maximum sequence length it supports. Thus, in this section, we compare our proposed design with the first two state-of-the-art solutions, both supported in DeepSpeed. 

\label{sec:setup}We choose six widely used LLM models with different sizes. For GPT-like models, we have 2.7B, 6.7B, 13B, and 30B. For Llama, we use the 8B and 70B models. For Megatron-SP, we follow its default parallel setting, which leverages tensor model parallel and sequence parallel. For our proposed FPDT, ZeRO-3 model parameter sharding is leveraged, this ensures fair comparison between Megatron-SP, DeepSpeed Ulysses, and FPDT. For all experiments, we set the batch size to 1 and enable the activation checkpoint with CPU offloading. Our goal is to fully exploit the long sequence handling capability of each parallel strategy. 

\begin{figure*}[t]
    \centering
        \begin{subfigure}{0.24\textwidth}
            \centering
            \includegraphics[width=\linewidth]{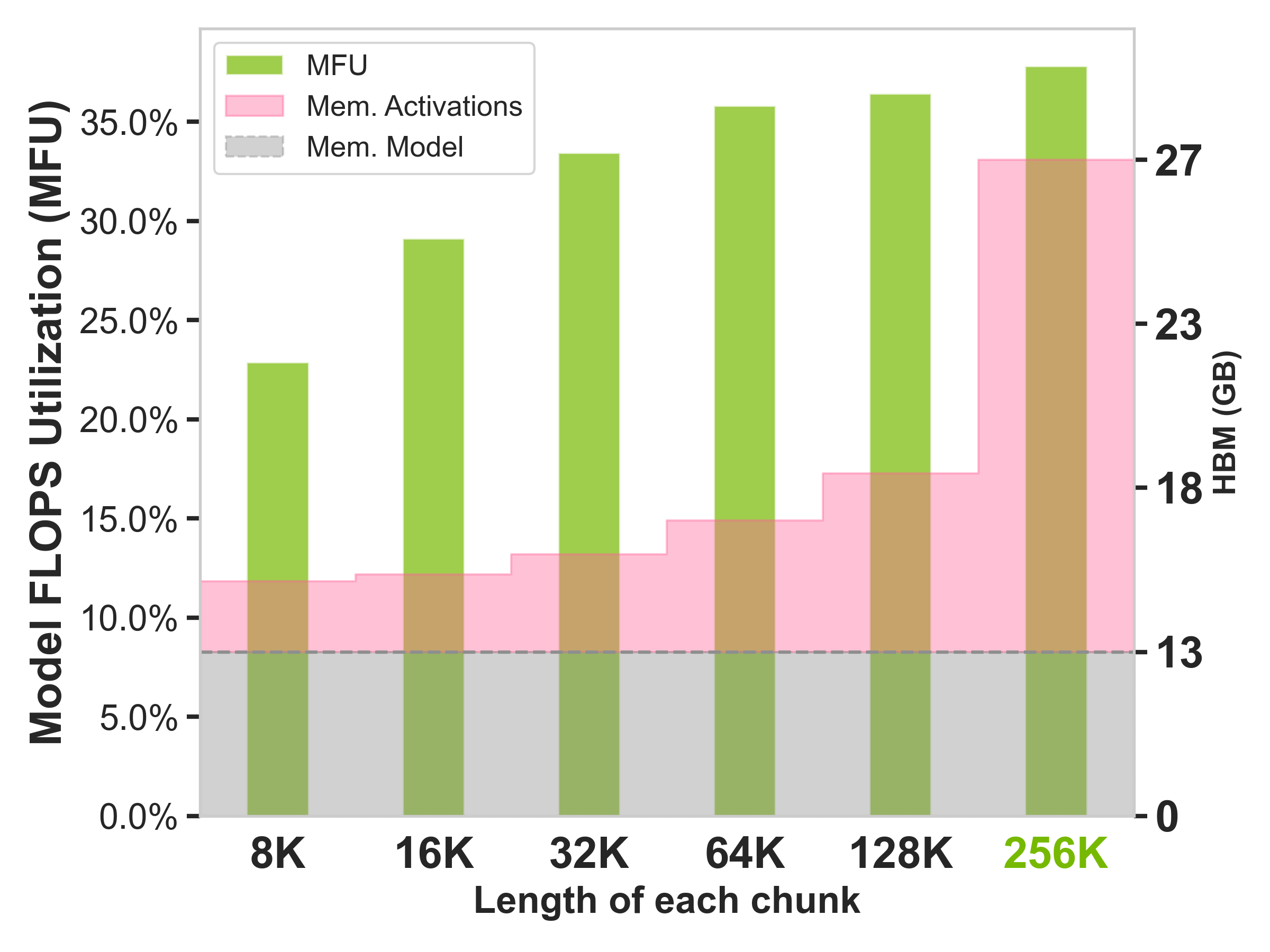}
            \caption{2.7B, p\&o. take 13.3GB.}
            \label{fig:2.7b_mem}
        \end{subfigure} 
        \begin{subfigure}{0.24\textwidth}
            \centering
            \includegraphics[width=\linewidth]{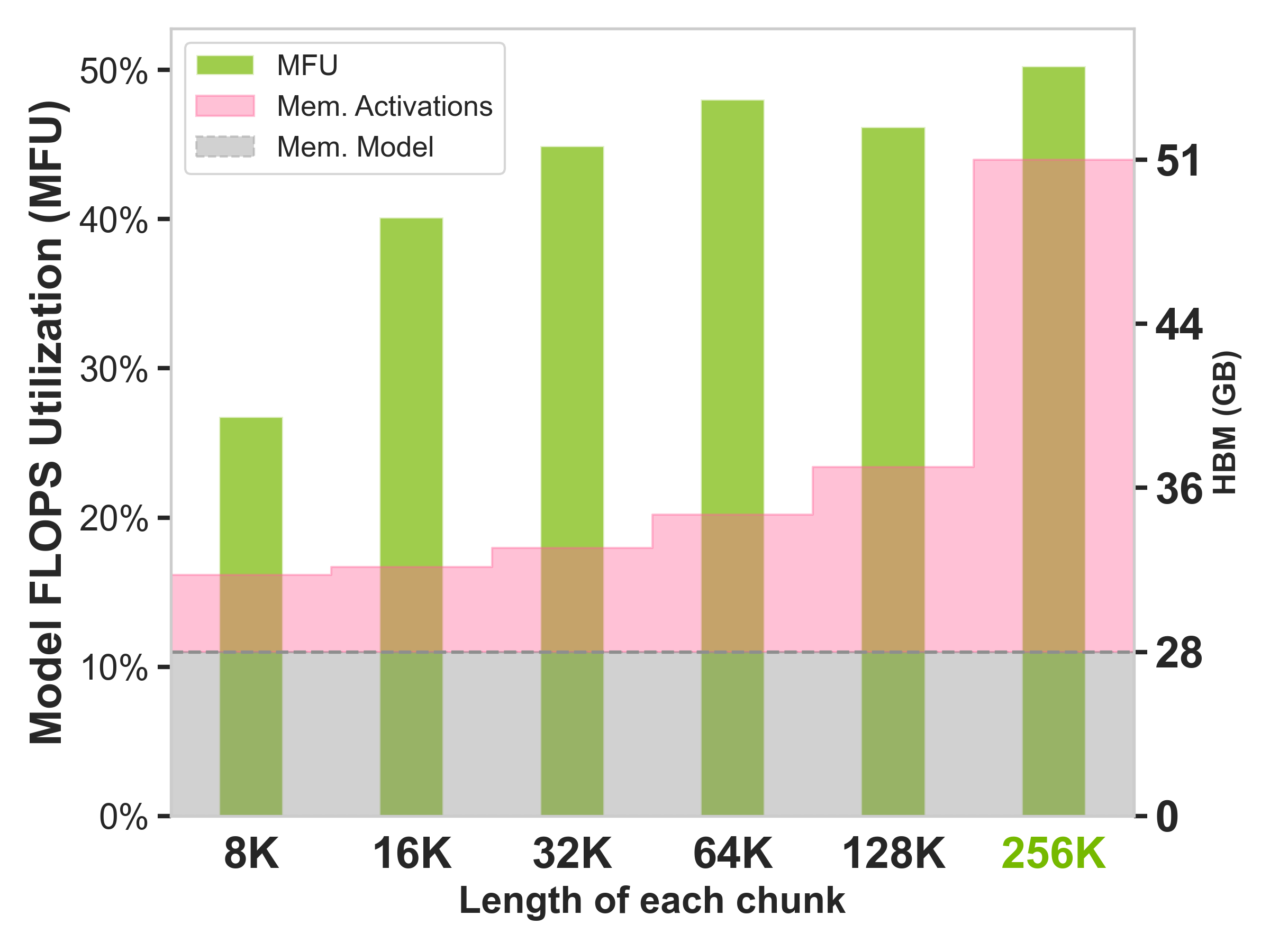}
            \caption{6.7B, p\&o. take 28.2GB.}
            \label{fig:6.7b_mem}
        \end{subfigure} 
        \begin{subfigure}{0.24\textwidth}
            \centering
            \includegraphics[width=\linewidth]{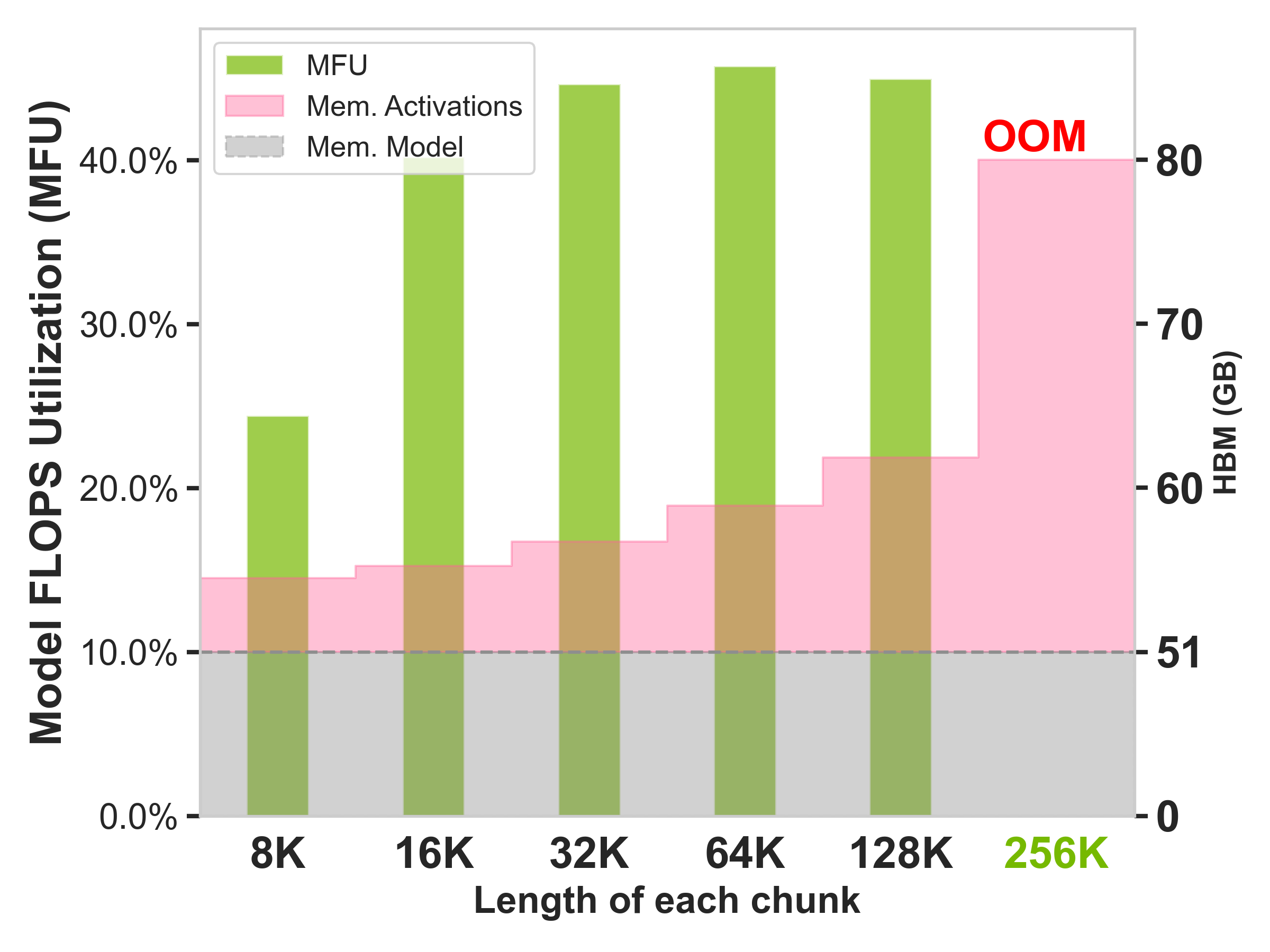}
            \caption{13B, p\&o. take 51.4GB.}
            \label{fig:13b_mem}
        \end{subfigure} 
        \begin{subfigure}{0.24\textwidth}
            \centering
            \includegraphics[width=\linewidth]{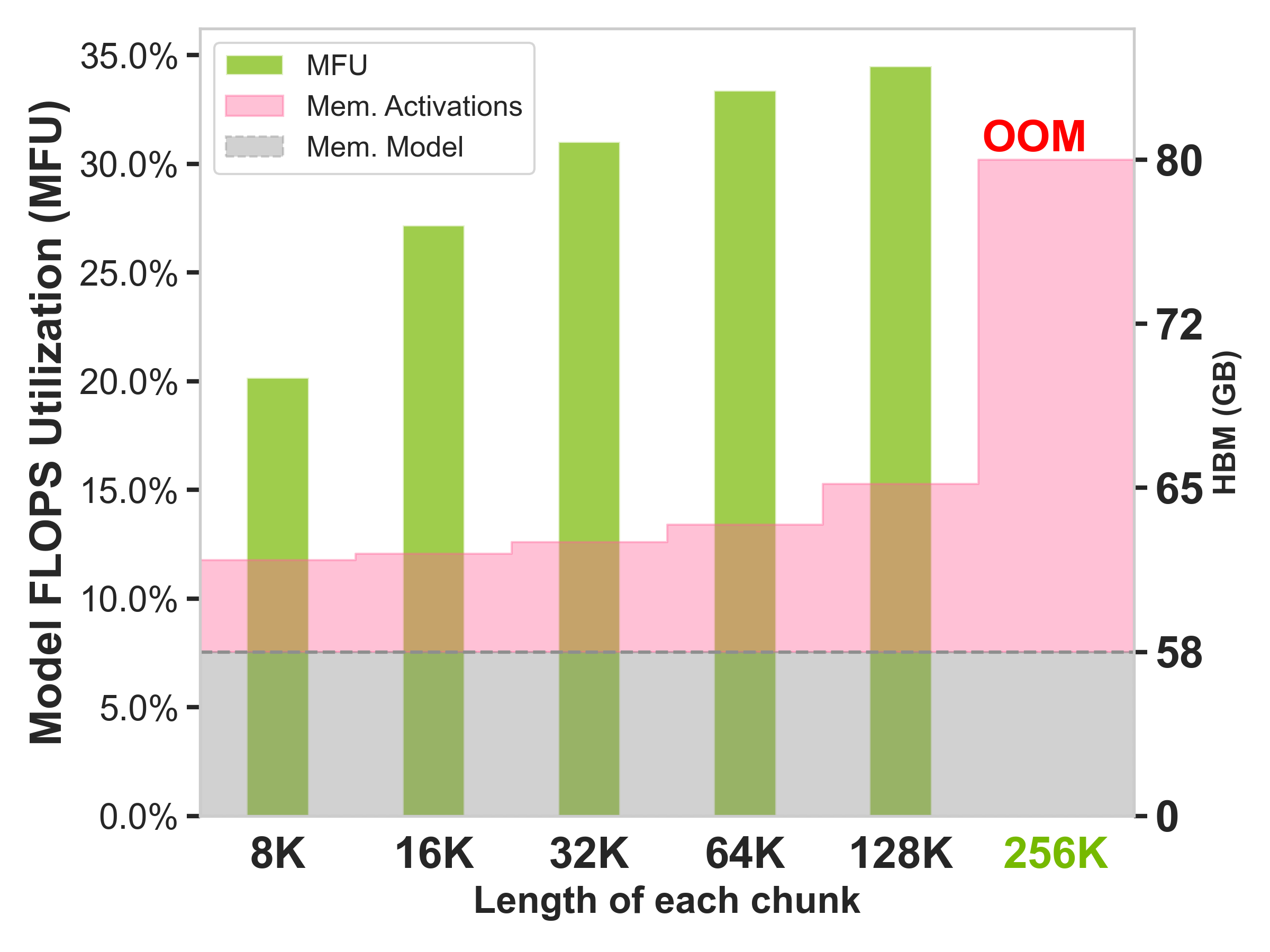}
            \caption{30B, p\&o. take 58GB.}
            \label{fig:13b_mem}
        \end{subfigure}
        \vspace{-0.5em}
    \caption{MFU and HBM memory consumed during training. The global sequence length is set to 256K, and we change the chunk size. We use 4 GPUs for 2.7B, 6.7B, and 13B models; 8 GPUs for 30B model. Gray areas represent the memory taken by model parameters and optimizer states (\textbf{p\&o.}). Pink areas represent the memory taken by activations.}
    \label{fig:mem}
\end{figure*}

Figure~\ref{fig:mfu} shows the end-to-end training performance. When running within one compute node, Megatron-SP and Ulysses exhibit similar hardware efficiency. For example, on the 2.7B GPT model, both methods have a maximum sequence length of 256K. For our proposed FPDT, with only chunking, we increase the sequence length by 8x longer, from 256K to 2M, without sacrificing performance. For the 6.7B GPT model, Megatron-SP and Ulysses 
support 128K and 256K sequence lengths. Our FPDT with chunking increases the bound to 512K while facing out-of-memory (OOM) issues when we keep increasing the sequence length. We then leverage the host memory. By offloading idle tokens from HBM, we support the 2M sequence length, which is 8x longer than the existing solutions. By properly managing the offloading, fetching, and computation, our FPDT with double buffer strategy reaches comparable hardware MFU as the non-offloading counterparts (i.e. FPDT w. chunking). Similar results are also shown in the 8B Llama 3 model training. As we increase the model's size, we scale it up to multiple nodes. We found that Ulysses is generally more efficient than Megatron-SP, as the latter's performance degrades severely when inter-node communication is included. For the 13B, 30B, and 70B models, we use 2, 4, and 8 nodes, respectively. While both Megatron-SP and Ulysses struggle with the memory spikes of activations, our FPDT works effectively by increasing the maximum sequence length to 2M while maintaining high MFU.

\subsection{Tradeoff on Sequence Chunk Size}
As discussed in ~\ref{sec:double_buffer}, choosing a proper chunk size can not only exploit the computing power of the hardware but also allow the data moving from host to device and from device to host to overlap by computation. In Figure~\ref{fig:mfu}, we use a default chunk size of 64K for all our FPDT-based methods. In this part, we will demonstrate why this is our preference. Figure~\ref{fig:mem} shows the usage of GPU HBM at different chunk lengths. We conduct the profiling on a single node with 4 GPUs, with a fixed global sequence length of 256K. We change the length of each sequence chunk and compare the corresponding memory footprint and MFU. The chunk size of 256K means that we do not use chunk, 
but just run the baseline Ulysses. 8K, 16K, 32K, 64K, and 128K of chunk size corresponds to 32, 16, 8, 4, and 2 chunks in total in our FPDT pipeline scheme. As shown in Figure~\ref{fig:2.7b_mem}, ~\ref{fig:6.7b_mem}, ~\ref{fig:13b_mem}, our FPDT can significantly reduce the activation memory footprint; for example, in the 2.7B model, we reduce the activation memory from 27GB to 18GB by splitting the sequence into two chunks. Despite that increasing the number of chunks can further reduce the memory footprint, we found that 64K is a sweet point where the latency of offloading and prefetching can be hidden by the computation. Less chunks means a shorter pipeline, which makes latencies of the first data preparation and the last computation more salient, as they cannot overlap with other operations. More chunks, however, make the latency of computation too short to hide that of data loading. Therefore, we choose 64K as it has an overall best-overlapped pipeline.

\begin{table*}[t]
    \centering
    {\fontfamily{ppl}\selectfont
        \begin{tabular}{c|cccccccc|ccc}
        \hline
        \multirow{13}{*}{\shortstack{8B Llama 3 \\ 8 GPUs}} & \multicolumn{8}{c}{Training strategies} & \multicolumn{3}{c}{Performance} \\
        & TP. & AC. & OC. & UL. & ZeRO-1 & ZeRO-2 & ZeRO-3 & FPDT & Max len. &  HBM. & MFU \\
        \cline{2-12}
         & \checkmark &  &  &  & & &   &  & 32K & 64.3G & 9.4\% \\
         & \checkmark & \checkmark &  &  &  & & &  & 128K & 61.2G & 19.4\% \\
         & \checkmark & \checkmark & \checkmark &  &  &  &&  & 512K  & 78.7G & 32.7\% \\
         &  &  &  & \checkmark & \checkmark & & &  & 64K & 58.9G & 15.3\% \\
         &  &  &  & \checkmark &  & \checkmark& &  & 64K & 54.5G & 15.3\% \\
         &  &  &  & \checkmark &  & & \checkmark&  & 64K & 52.3G & 21.0\% \\
         &  & \checkmark  & \checkmark & \checkmark & \checkmark & & &  & 512K & 65.5G & 46.8\% \\
         &  & \checkmark  & \checkmark & \checkmark & & \checkmark & &  & 512K & 65.5G & 46.8\% \\
         &  & \checkmark  & \checkmark  &\checkmark & & & \checkmark &  & 512K & 60.1G & 47.2\% \\
         &  & \checkmark  & \checkmark  & & & & \checkmark & \ding{72} & \textbf{4M} & 68.0G & \textbf{55.7\%} \\
        \hline
        \end{tabular}
    }
\caption{A comprehensive analysis on long-context LLM training with different training techniques. \textbf{TP.} denotes tensor parallel. \textbf{AC.} denotes activation checkpoint. \textbf{OC.} denotes activation checkpoint with CPU offloading. \textbf{UL.} stands for Ulysses. \textbf{FPDT} is our proposed Fully Pipelined Distributed Transformer.}
\label{tab:comprehensive_strategies}
\vspace{-1em}
\end{table*}

\subsection{Chunk Granularity}

\begin{figure}[H]
    \centering
    \begin{subfigure}{\linewidth}
        \centering
        \includegraphics[width=0.95\linewidth]{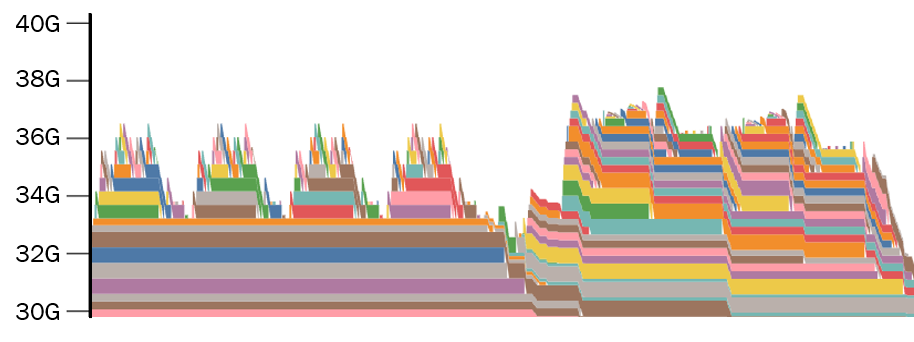}
        \vspace{-0.5em}
        \caption{2 chunks in attention, 4 chunks in FFN}
    \end{subfigure}\\
    \begin{subfigure}{\linewidth}
        \centering
        \includegraphics[width=0.95\linewidth]{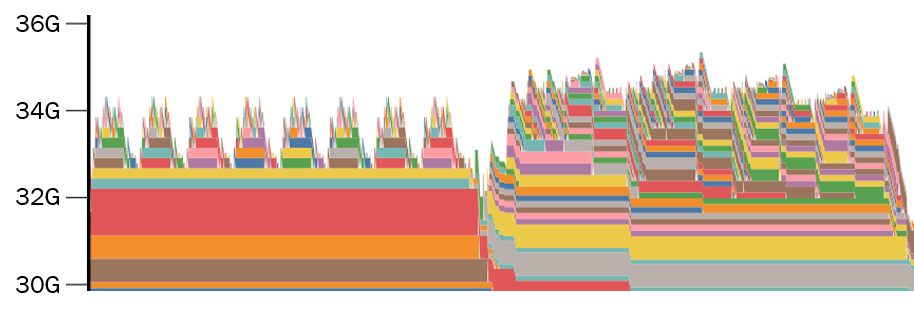}
        \vspace{-0.5em}
        \caption{4 chunks in attention, 8 chunks in FFN}
    \end{subfigure}\\
    \begin{subfigure}{\linewidth}
        \centering
        \includegraphics[width=0.95\linewidth]{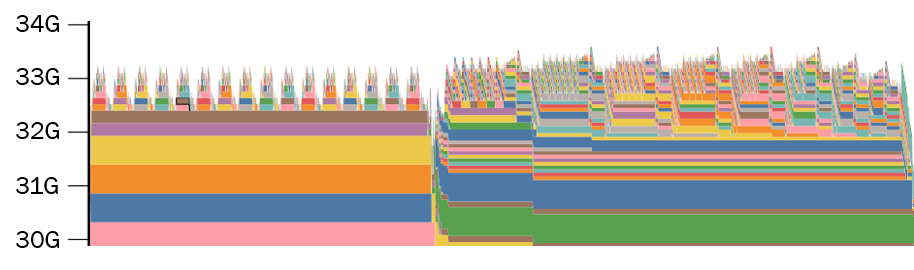}
        \vspace{-0.5em}
        \caption{8 chunks in attention, 16 chunks in FFN}
    \end{subfigure}
    \caption{Memory footprint in the backward pass of a Transformer block in Llama 8B. We first calculate the gradients in FFN, then the attention. FFN uses 2x chunks of attention.}
    \label{fig:mem_footprint}
\end{figure}

As we analyzed in table~\ref{tab:memory_transformer}, in forward and backward passes, attention operation and FFN can incur different amounts of intermediate buffers, therefore, different chunking strategies need to be applied. The chunking and offloading strategies of the attention part have been introduced in ~\ref{sec:double_buffer}. For FFN, however, we cannot easily leverage offloading to reduce GPU memory consumption without significantly sacrificing hardware efficiency. For token-wise operations such FFN, the complexity of computation increases linearly, i.e. $O(N)$. In this case, $\text{F(N)} = \Theta(\text{G(N)})$, where $F$ and $G$ are the complexities of compute and memory, respectively. Considering the high throughput of GPU compute cores, the latency of offloading and prefetching can never be overlapped by computation, thus, for FFN, we don't use offloading. Figure~\ref{fig:mem_footprint} are the memory profiles created by the PyTorch profiler (note that PyTorch randomly picks colors in each profile), we found that for the GPT and Llama models, setting the number of chunks in the FFN to be twice that of the attention is sufficient to ensure that the attention part strictly binds the memory footprint. 

Also noteworthy is that since we do not offload chunks in FFN, as long as the size of each chunk is not too small, the overall training throughput would not be significantly affected. We also want to emphasize that another memory spike during training is found in the final calculation of softmax and cross-entropy loss. As the vocabulary size is much larger than the model's hidden dimension, and the operation usually requires a \texttt{Float32} data type, the last linear project would lead to the out-of-memory issue. However, this part can be solved by chunking as well, and since it is at the end of the forward pass, the number of chunks used in this part is trivial to the overall performance. Thus, we suggest that setting it to $\frac{vocab\_size}{hidden\_dim}\times2$, would solve the memory spike.
\vspace{-1em}
    \vspace{-1em}

\subsection{Training strategies in long-context LLM}
Though we discussed our experiments setup in~\ref{sec:setup}, we would like to know how each strategy contributes to the long-context LLM training. Tensor parallel is widely used in distributed model training. It allows each GPU to only keep a slice of the tensor along the hidden dimension, hence also parallelizing the computation. However, it cannot reduce the memory footprint of intermediate buffers, since the GEMM between tensor [$N, B, C$] and weight [$\hat{C}, C$] generates an intermediate buffer [$N, B, \hat{C}$] regardless of $C$. Activation checkpoint (\textbf{AC.}) is also a commonly used strategy in large model training, as it can significantly reduce the GPU memory pressure for models with many layers. Activation checkpoint with offloading (\textbf{OC.}) is a technique introduced in DeepSpeed, which allows the checkpointed activation to be moved to host memory. Both \textbf{AC.} and \textbf{OC.} play critical roles in enlarging the sequence length, i.e., from 32K to 128K to 512K. For Ulysses, it can seamlessly work with the ZeRO family, which reduces a decent amount of memory footprint, i.e., from 58.9G to 52.3G. However, ZeRO doesn't mitigate the memory requirement for long-context LLM training as it reduces only memory usage related to the model parameters. 
Finally, our proposed FPDT, together with activation checkpointing, activation offloading, and ZeRO-3, remarkably increases the maximum sequence length by 8x, from 512K to 4M, on an 8B Llama 3 with 8 GPUs, while achieving extreme hardware efficiency.

\subsection{Sparse attention}

Sparse attention is crucial for improving the efficiency of long-context large language models (LLMs) by reducing the computational and memory costs of standard attention mechanisms, which scale quadratically with sequence length.
 By focusing only on the most relevant tokens, sparse attention significantly speeds up inference while preserving model accuracy. In Table~\ref{tab:sparsity}, we use the block sparse attention to test the training MFU with our FPDT design. Note that, with sparse attention, only part of the tokens in key and value will be fetched from the host memory, while the query will always be the entire sequence. As we discussed in Figure~\ref{fig:avg_time}, this may leads to an overall shift towards left, where the latency of fetching cannot be perfectly overlapped by computation, which explains the slightly lower MFU when sparsity is high.
\begin{table}[H]
    \centering
    {\fontfamily{ppl}\selectfont
        \begin{tabular}{c|cccccc}
        \hline
        \multirow{2}{*}{\shortstack{Models}}& \multicolumn{6}{c}{Attention Sparsity}\\
        & 0.5 & 0.4 & 0.3 & 0.2 & 0.1 & 0.0 \\
        \cline{1-7}
        2.7B & 41.7 & 38.6 & 38.8 & 39.0 & 38.9 & 38.4 \\
        8B & 40.6 & 42.0 & 44.0 & 46.4 & 46.5 & 47.6 \\
        13B & 40.7 & 42.8 & 44.2 & 45.4 & 46.4 & 46.1 \\
        \hline
        \end{tabular}
    }
\caption{MFU(\%) at different attention sparsity. A higher sparsity means less KV tokens are fetched for computation. We use 1 GPU for 2.7B model, and 4 GPUs for 8B and 13B models. Chunk size is 64K and global sequence length is 256K. 0.0 means full attention.}
\vspace{-1em}
\label{tab:sparsity}
\end{table}
\section{Future work}
This paper focuses on alleviating the memory constraints that are brought by the activations and intermediate buffers in long-sequence LLM training. We conduct experiments with the ZeRO-3 technique. However, we noticed that PyTorch here can also incur a high memory spike when it reduces the gradients across all GPUs. In certain cases, this memory spike can be more significant than the activation's memory spikes, which becomes a bottleneck in keeping increasing sequence length. We will investigate this and welcome researchers in this field to advance long-sequence LLM training together.
\section{Conclusion}
In this paper, we present the Fully Pipelined Distributed Transformer (FPDT), for efficiently training long-sequence LLMs within resource-constrained environment. Our proposed method leverages advanced sequence parallelism and ZeRO-3, largely reducing the GPU resource required for million-level sequence training. With our elaborately designed overlapping scheme, training 2.7B to 70B LLMs on up to 4M token sequence with FPDT reaches over 55\% MFU. Our method can also be applied to LLMs with Transformer-like blocks with little re-configuration. We believe our work can benefit the large community in exploring LLMs' capability in long context scenarios.

\bibliographystyle{mlsys2025}
\bibliography{main}

\appendix
\section{Convergence Evaluation}
\begin{figure}[H]
    \centering
    \includegraphics[width=0.95\linewidth]{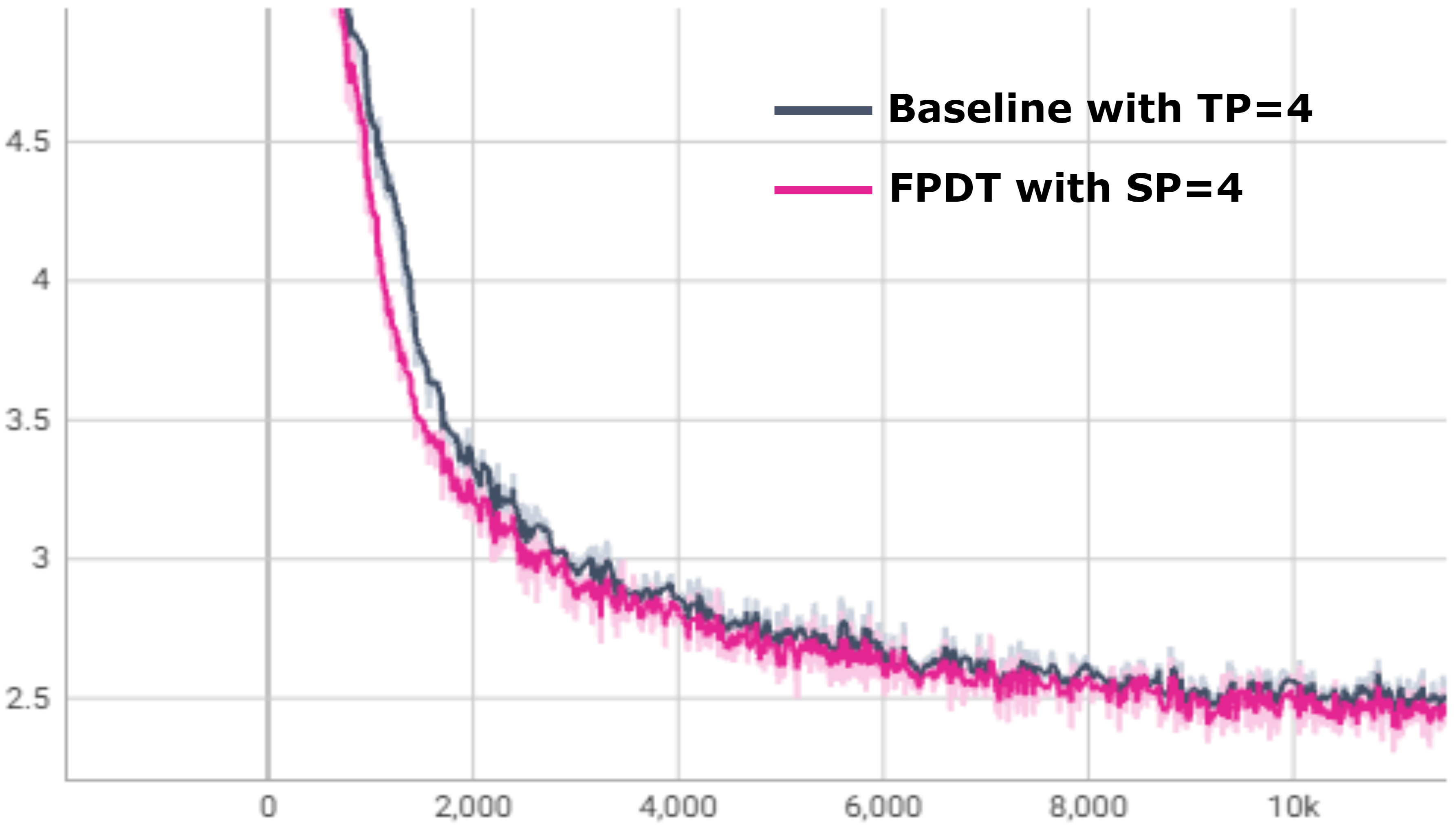}
    \vspace{-0.5em}
    \caption{Loss curve in pretraining GPT models with 4 GPUs.}
    \label{fig:convergence}
\end{figure}
Figure~\ref{fig:convergence} shows the convergence of the baseline GPT model that leverages tensor parallelism on 4 GPUs, with a batch size of 256 and ZeRO-1 enabled, and our FPDT with and without offloading. Our proposed FPDT is a pure system optimization technique that enables the training of ultra-long sequence Transformer models without affecting the quality of the trained models.

\section{Artifact Evaluation}

\subsection*{1. Artifact Abstract}
\begin{itemize}
    \item \textbf{Purpose:} This artifact reproduces the results illustrated in Figure~\ref{fig:mfu} and Figure~\ref{fig:mem}, showing end-to-end training MFU on various models and sequence lengths.
    \item \textbf{Scope:} We have integrated our code into the open-sourced DeepSpeed and Megatron-DeepSpeed repositories. Users can follow the instructions in these repositories to set up the environment and run the experiments.
    \item \textbf{Requirements:} The artifact requires machines equipped with PCIe 4.0, NVIDIA A100 80GB GPUs with NVLINK, and high-speed interconnects such as InfiniBand/RoCE.
    \item \textbf{Access:} The artifact is available at \href{https://github.com/deepspeedai/DeepSpeed/blob/master/blogs/ulysses-offload/README.md}{DeepSpeed Ulysses-Offload}.
\end{itemize}

\subsection*{2. Artifact Contents}
\begin{itemize}
    \item \textbf{Data:} Datasets such as BookCorpus, Pile, etc.
    \item \textbf{Models:} GPT and LLaMA models, with sizes ranging from 2.7B to 70B.
    \item \textbf{Automation Scripts:} The distributed training script is provided \href{https://github.com/deepspeedai/Megatron-DeepSpeed/blob/main/examples_deepspeed/sequence_parallel/ds_pretrain_gpt_6.7B_fpdt_32k.sh}{here}.
    \item \textbf{Documentation:} Users can follow the tutorial available \href{https://www.deepspeed.ai/tutorials/ulysses-offload/}{here}.
\end{itemize}

\subsection*{3. Dependencies and Environment}
\begin{itemize}
    \item \textbf{Software:} PyTorch 2.5.1, CUDA 12.1, flash-attn 2.6.3, DeepSpeed, and Megatron-DeepSpeed.
    \item \textbf{Hardware:} NVIDIA A100 80GB GPUs with NVLINK, and InfiniBand interconnects.
\end{itemize}

\subsection*{4. Setup and Installation}
\begin{enumerate}
    \item \textbf{Installation:} Users can follow the standard installation procedures for DeepSpeed and Megatron-DeepSpeed. Note that for kernel fusion, Megatron-DeepSpeed requires NVIDIA APEX. Additionally, Flash-Attention is needed for efficient attention computation.
\end{enumerate}

\subsection*{5. Experimental Workflow}
\begin{enumerate}
    \item \textbf{Running the Experiments:} Follow the instructions in the provided automation script to execute the distributed training.
    \item \textbf{Validation:} The script logs throughput and performance metrics. Users can enable or disable FPDT and offload features to reproduce results shown in Figure~\ref{fig:mfu}.
    \item \textbf{Ablation Study:} Modify the chunk size parameter (default is 65536) in the script to replicate the memory results from Figure~\ref{fig:mem}.
\end{enumerate}

\subsection*{6. Validation and Reproducibility Checklist}
\begin{itemize}
    \item \textbf{Correctness Check:} Unit tests for numerical correctness are provided \href{https://github.com/deepspeedai/DeepSpeed/blob/master/tests/unit/sequence_parallelism/test_ulysses.py}{here}, comparing the baseline tensor-parallel training outputs with our framework.
    \item \textbf{Throughput Check:} Running the distributed training script will output throughput metrics to validate the performance.
\end{itemize}

\subsection*{7. Additional Notes and Corrections}
\begin{itemize}
    \item \textbf{Known Issues:} The standard output of DeepSpeed and Megatron-DeepSpeed is based on FLOPs. For auto-regressive models, since the attention computation only works on half of the attention matrix, a further conversion is required as the FLOPs calculation in the repository is based on the full attention matrix. We correct it and convert it to MFU.
    \item \textbf{Future Work:} Our framework currently supports all power-of-two sequence lengths, as demonstrated in all experiments in our paper. Support for arbitrary sequence lengths is in progress and will be available soon.
\end{itemize}
\end{document}